\documentclass[conference]{IEEEtran}

\usepackage{algorithm}
\usepackage{algpseudocode}
\usepackage{amsmath}
\usepackage{graphicx}
\usepackage{multirow}
\usepackage{booktabs}
\usepackage{xcolor}
\usepackage{subcaption}
\usepackage{siunitx}
\usepackage{tabularx}
\usepackage{adjustbox}
\usepackage{url}
\usepackage{enumitem}
\usepackage{hyperref}
\usepackage{tikz}

\usepackage[T1]{fontenc} \usepackage{cite}

\newcommand{\MVDRAM}{MVDRAM}
\newcommand{\Fig}[1]{Fig.~\ref{#1}}
\newcommand{\arxiv}[1]{#1}

\def\gemvLatencyOverCpu{7.29$\times$}
\def\gemvLatencyOverGpu{8.55$\times$}
\def\gemvEnergyOverCpu{30.5$\times$}
\def\gemvEnergyOverGpu{8.87$\times$}
\def\llmTwoThroughputOverCpu{2.18$\times$}
\def\llmTwoThroughputOverGpu{3.33$\times$}
\def\llmFourThroughputOverCpu{1.31$\times$}
\def\llmTwoEnergyOverCpu{3.04$\times$}
\def\llmTwoEnergyOverGpu{1.83$\times$}
\def\llmFourEnergyOverCpu{2.35$\times$}

\newcommand{\pudUnmodifiedRef}{
    gaoComputeDRAMInMemoryCompute2019,
    olgunQUACTRNGHighThroughputTrue2021,
    gaoFracDRAMFractionalValues2022,
    yukselFunctionallyCompleteBooleanLogic2024,
    yukselSimultaneousManyRowActivation2024}
\newcommand{\pudUnmodifiedCite}{\cite{\pudUnmodifiedRef}}
\newcommand{\pudModifiedSmallRef}{seshadriRowCloneFastEnergyefficient2013,seshadriAmbitInmemoryAccelerator2017,aliInMemoryLowCostBitSerial2020,ferreiraPLUToEnablingMassively2022}
\newcommand{\pudModifiedSmallCite}{\cite{\pudModifiedSmallRef}}
\newcommand{\pudModifiedLargeRef}{liDRISADRAMbasedReconfigurable2017,dengDrAccDRAMBased2018,dengLAccExploitingLookup2019,lenjaniFulcrumSimplifiedControl2020}
\newcommand{\pudModifiedLargeCite}{\cite{\pudModifiedLargeRef}}
\newcommand{\pudModifiedFrameRef}{hajinazarSIMDRAMFrameworkBitserial2021,oliveiraMIMDRAMEndtoEndProcessingUsingDRAM2024}
\newcommand{\pudModifiedFrameCite}{\cite{\pudModifiedFrameRef}}
\newcommand{\pudModifiedCite}{\cite{\pudModifiedSmallRef,\pudModifiedLargeRef,\pudModifiedFrameRef}}
\newcommand{\pndRef}{
    kwon25420nm6GB2021,
    heoNeuPIMsNPUPIMHeterogeneous2024,
    parkAttAccUnleashingPower2024,
    seoIANUSIntegratedAccelerator2024,
    liSpecPIMAcceleratingSpeculative2024,
    kwonLoLPIMLongContextLLM2025,
    hePAPIExploitingDynamic2025}
\newcommand{\pndCite}{\cite{\pndRef}}

\newcommand{\PRE}{\texttt{PRE}}
\newcommand{\ACT}{\texttt{ACT}}

\newcommand{\AND}{\texttt{AND}}

\newcommand{\NOT}{\texttt{NOT}}

\newcommand{\RowCopy}{\texttt{RowCopy}}
\newcommand{\MAJ}[1]{\texttt{MAJ#1}}

\graphicspath{{fig/}}  \newcommand*\circled[1]{\tikz[baseline=(char.base)]{\node[circle,draw,inner sep=0.5pt] (char) {#1};}}

 \IEEEoverridecommandlockouts

\begin{document}

\title{\MVDRAM{}: Enabling GeMV Execution in Unmodified DRAM for Low-Bit LLM Acceleration}

\makeatletter
\newcommand{\linebreakand}{\end{@IEEEauthorhalign}
  \hfill\mbox{}\par
  \mbox{}\hfill\begin{@IEEEauthorhalign}
}
\makeatother

\author{

    \IEEEauthorblockN{Tatsuya Kubo\IEEEauthorrefmark{1}}
    \IEEEauthorblockA{
    The University of Tokyo}
    \and
    \IEEEauthorblockN{Daichi Tokuda}
    \IEEEauthorblockA{
    The University of Tokyo}
    \and
    \IEEEauthorblockN{Tomoya Nagatani}
    \IEEEauthorblockA{
    The University of Tokyo}
    \and
    \IEEEauthorblockN{Masayuki Usui}
    \IEEEauthorblockA{
    The University of Tokyo}
    \linebreakand
    \IEEEauthorblockN{Lei Qu}
    \IEEEauthorblockA{Microsoft Research}
    \and
    \IEEEauthorblockN{Ting Cao\IEEEauthorrefmark{1}}
    \IEEEauthorblockA{Institute for AI Industry Research (AIR)\\Tsinghua University}
    \and
    \IEEEauthorblockN{Shinya Takamaeda-Yamazaki}
    \IEEEauthorblockA{
    The University of Tokyo}

    \thanks{\IEEEauthorrefmark{1}Corresponding authors: Tatsuya Kubo \texttt{<}tatsuya.kubo@is.s.u-tokyo.ac.jp\texttt{>} and Ting Cao \texttt{<}tingcao@mail.tsinghua.edu.cn\texttt{>}.}
}
 
\maketitle
\thispagestyle{plain}
\pagestyle{plain}

\begin{abstract}
General matrix-vector multiplication (GeMV) remains a critical latency bottleneck in large language model (LLM) inference, even with quantized low-bit models.
Processing-Using-DRAM (PUD), an analog in-DRAM computing technique, has the potential to repurpose on-device DRAM as a GeMV engine, offering additional high-throughput processing capabilities to widespread consumer devices without DRAM modifications.
However, applying PUD to GeMV operations in the LLM inference pipeline incurs significant overheads \textit{before} and \textit{after} in-DRAM computation, diminishing the benefits of its high-throughput processing capabilities.

This paper presents \MVDRAM{}, the first practical system to accelerate GeMV operations for low-bit LLM inference using unmodified DRAM.
By leveraging the data sharing patterns and mathematical linearity in GeMV operations, \MVDRAM{} orchestrates the processor and DRAM to eliminate the costs associated with pre-arranging inputs and bit-transposition of outputs required in conventional PUD approaches.
Our experimental evaluation with four DDR4 DRAM modules shows that \MVDRAM{} achieves comparable or even better inference speed than the processor-based implementation for GeMV operations in low-bit (under 4-bit) LLM.
In particular, \MVDRAM{} achieves up to \gemvLatencyOverCpu{} speedup and \gemvEnergyOverCpu{} energy efficiency for low-bit GeMV operations.
For end-to-end LLM inference, \MVDRAM{} achieves \llmTwoThroughputOverCpu{} and \llmFourThroughputOverCpu{} throughput improvements, along with \llmTwoEnergyOverCpu{} and \llmFourEnergyOverCpu{} energy efficiency, for 2-bit and 4-bit quantized low-bit models, respectively.
\arxiv{
    \MVDRAM{} has the potential to redefine the AI hardware landscape by demonstrating the feasibility of standard DRAM as an LLM accelerator.
}
\end{abstract}
 
\section{Introduction}
Large language models (LLMs) are increasingly deployed on consumer devices as integral underlying system components, such as the on-device 2/4-bit 3B Apple foundation model for Apple's iOS~\cite{gunterAppleIntelligenceFoundation2024}, the 4-bit 3.82B Phi Silica for Windows~\cite{pradeep2024phi}, and 4-bit 3.35B Gemini Nano for Google's Android~\cite{teamGeminiFamilyHighly2024}.
These models generate intensive DRAM accesses during inference due to the dominant large-scale general matrix-vector multiplication (GeMV) operations, for key-value (KV) cache and single-batch feed-forward network (FFN) calculation~\cite{kwonLoLPIMLongContextLLM2025,parkAttAccUnleashingPower2024,hePAPIExploitingDynamic2025,seoIANUSIntegratedAccelerator2024,heoNeuPIMsNPUPIMHeterogeneous2024}.
This problem is even more pronounced in current chain-of-thought (CoT) reasoning models with a long context~\cite{chengCompressedChainThought2024}.
This GeMV bottleneck limits token generation performance and energy efficiency, especially in consumer devices with restricted memory bandwidth and power budgets.

To address memory constraints in LLM deployment, low-bit quantization has become the de facto technique.
This approach represents numerical values of activations and weights in a lower number of bits.
Recent techniques achieve minimal accuracy loss using $4$-bit representations~\cite{ashkboosQuaRotOutlierFree4Bit2024,zhao2024atom,elangovanBCQBlockClustered2025} while more aggressive methods employing $3$-bit, $2$-bit, and even $1$-bit precision continue to emerge~\cite{wangBitNetScaling1bit2023,duBitDistillerUnleashingPotential2024,chee2023quip,tsengQuIPEvenBetter2024,vptq,linQServeW4A8KV4Quantization2024,xuOneBitExtremelyLowbit2024}.
These low-bit quantization techniques effectively reduce the memory footprint and computational complexity~\cite{wuUnderstandingINT4Quantization2023}.

\textit{Processing-Using-DRAM} (PUD)\footnote{
    This paper follows the definition in~\cite{mutluMemoryCentricComputingRecent2024}.
    DRAM-based Processing-in-Memory (PiM) includes: (1)  Processing-near-DRAM (PnD):  computation logic is added near the memory arrays; (2) Processing-Using-DRAM (PUD): exploit the analog operational properties of the memory.
} has emerged as an innovative technique that leverages the analog operational characteristics of DRAM to execute highly parallel bit-serial computations directly within memory arrays.
Unlike approaches that demand specialized memory circuits and chips~\pudModifiedCite{}, some of existing PUD techniques demonstrate the potential to enable in-memory computation using commercial off-the-shelf DRAM without DRAM hardware modifications~\pudUnmodifiedCite{}.
By intentionally issuing DRAM commands that violate manufacturer-specified timing parameters, PUD with unmodified DRAM provides two fundamental operations: \RowCopy{}, which transfers data between rows by exploiting incomplete bitline precharging, and \textit{majority-of-X} (\MAJ{X}), which computes the majority voting of $X$ cells connected to the same bitline.
These operations can be performed simultaneously across all columns in a bank, offering massive parallelism up to 65,536 bitwise operations in parallel.

The meeting of massively parallel bitwise operations in PUD and large low-bit GeMV in quantized LLMs inspires us to raise the question: \textit{can off-the-shelf on-device DRAM serves not only as model storage but also as an inference accelerator for GeMV operations?}
The ability to leverage computational capabilities from on-device DRAM offers a practical path for realizing LLM inference on resource-constrained devices without requiring specialized hardware. Also, it can potentially release the precious on-device computing, energy and memory bandwidth for other applications. 

However, applying current PUD techniques~\cite{gaoComputeDRAMInMemoryCompute2019,yukselFunctionallyCompleteBooleanLogic2024,hajinazarSIMDRAMFrameworkBitserial2021,oliveiraMIMDRAMEndtoEndProcessingUsingDRAM2024} to GeMV operations introduces too much overheads to be practically used in LLM inference, diminishing the benefits of PUD's high-throughput processing capabilities.
These overheads stem from the fundermental challenge of PUD computation mechanism: \textit{the limitation of column-to-column data movement}, which is the inability of PUD to move data across different columns in a DRAM subarray.
When applying PUD to GeMV operations in LLM inference, this limitation forces existing methods to have two issues:
(1) For a GeMV operation with a matrix of size $M \times N$ and an input vector of size $N$, the processor has to pre-arrange the input vector to memory by $M$ times so that it can execute $M$ parallel multiplication-and-accumulation (MAC) operations, introducing memory writing overheads;
(2) Conventional methods assign one column per computation, leading to lower utilization of PUD's parallelism and additional power overhead due to the processor's bit-transposition during data placement and retrieval from PUD.

This paper presents \MVDRAM{}, the first system to realize GeMV operations for end-to-end low-bit LLM inference using unmodified DRAM.
In contrast to approaches that enhance DRAM's computational capabilities through circuit modifications, \MVDRAM{} overcomes the limitations of PUD with unmodified DRAM through processor-DRAM co-design.

\MVDRAM{} addresses the challenges of GeMV acceleration through two novel techniques: \textit{on-the-fly vector encoding} and \textit{horizontal matrix layout}.
(1) The on-the-fly vector encoding technique dynamically generates PUD operation sequences based on the activation vector's bit values, eliminating the latency overheads associated with pre-arranging inputs and enabling an optimization for the sparsity of the activation vector.
(2) The horizontal matrix layout addresses the capacity and power overheads of conventional PUD's vertical layouts by exploiting the mathematical linearity of matrix-vector multiplication.
By decomposing MAC operations with respect to matrix bits and organizing the matrix elements in a row-wise manner, this approach enhances the utilization of PUD's parallelism and enables bit-transposition-free output aggregation through standard DRAM row access patterns.

In summary, this paper offers the following contributions:
\begin{itemize}[leftmargin=*]
    \item
        We present \MVDRAM{}, the first system to realize GeMV operations for end-to-end low-bit LLM inference using unmodified DRAM.
    \item
        We propose on-the-fly vector encoding, which directly encodes activation vector values into PUD operation sequences, eliminating the overheads associated with pre-arranging inputs and enabling efficient sparse activation optimization.
    \item
        We introduce horizontal matrix layout, which leverages the mathematical linearity of matrix-vector multiplication, enhancing the utilization of PUD's parallelism and enabling output aggregation without bit-transposition.
    \item
        We implement and evaluate \MVDRAM{} on a real system with four DDR4 DRAM modules, demonstrating comparable or even better inference speed than the processor-based implementation for GeMV operations in low-bit (under 4-bit) LLM.
        In particular, \MVDRAM{} achieves up to \gemvLatencyOverCpu{} speedup and \gemvEnergyOverCpu{} energy efficiency for low-bit GeMV operations, and \llmTwoThroughputOverCpu{} and \llmFourThroughputOverCpu{} throughput improvements, along with \llmTwoEnergyOverCpu{} and \llmFourEnergyOverCpu{} energy efficiency, for 2-bit and 4-bit quantized low-bit models, respectively.
\end{itemize}

\arxiv{
\MVDRAM{} represents a shift in how AI computations can be performed.
By transforming DRAM into an LLM accelerator through simple memory controller modifications, \MVDRAM{} reduces reliance for dedicated accelerators, enabling low-cost, power-efficient, and scalable AI deployment, even on mobile and wearable devices.
\MVDRAM{} has the potential to reshape the AI hardware landscape, making high-performance inference more accessible than ever before.
} \section{Background}

\subsection{LLM Inference and Low-Bit GeMV}
Large language models (LLMs) consist of multiple transformer-based decoder layers, with each layer containing attention mechanisms and feed-forward networks (FFNs).
These components involve various computational operations, including general matrix-vector multiplication (GeMV), softmax, and layer normalization.

Among the computational operations in LLM inference, GeMV is the main cost of LLM inference~\cite{kwonLoLPIMLongContextLLM2025,parkAttAccUnleashingPower2024,hePAPIExploitingDynamic2025,seoIANUSIntegratedAccelerator2024,heoNeuPIMsNPUPIMHeterogeneous2024}, requiring intensive memory access for key-value cache and single-batch FFN calculation.
This problem is exacerbated by recent reasoning models~\cite{deepseek-aiDeepSeekR1IncentivizingReasoning2025}, which can generate thousands of tokens during the chain-of-thought process.
Due to the low data reuse, a series of GeMV operations results in significant latency and energy overheads.

To address the memory bottleneck, low-bit quantization has become the de facto technique for deploying LLMs.
This technique represents the weights and activations with a lower number of bits.
For instance, the QuaRot~\cite{ashkboosQuaRotOutlierFree4Bit2024}, Atom~\cite{zhao2024atom} and BCQ~\cite{elangovanBCQBlockClustered2025} introduce $4$-bit weight and $4$-bit activation quantization with minimal accuracy loss.
Beyond $4$-bit quantization, models utilizing $3$-bit, $2$-bit, and even $1$-bit precision are emerging~\cite{wangBitNetScaling1bit2023,duBitDistillerUnleashingPotential2024,chee2023quip,tsengQuIPEvenBetter2024,vptq,linQServeW4A8KV4Quantization2024,xuOneBitExtremelyLowbit2024}. 
These low-bit quantization techniques can reduce the memory footprint and computational complexity~\cite{wuUnderstandingINT4Quantization2023}.
Consequently, recent system innovations shift traditional data-centric computations toward bit-wise operations, where low-bit multiplication is implemented as the sum of partial products for each bit~\cite{weiTMACCPURenaissance2024,parkLUTGEMMQuantizedMatrix2024,moLUTTensorCore2024}.

\begin{figure}[t]
    \centering
    \includegraphics[width=\columnwidth]{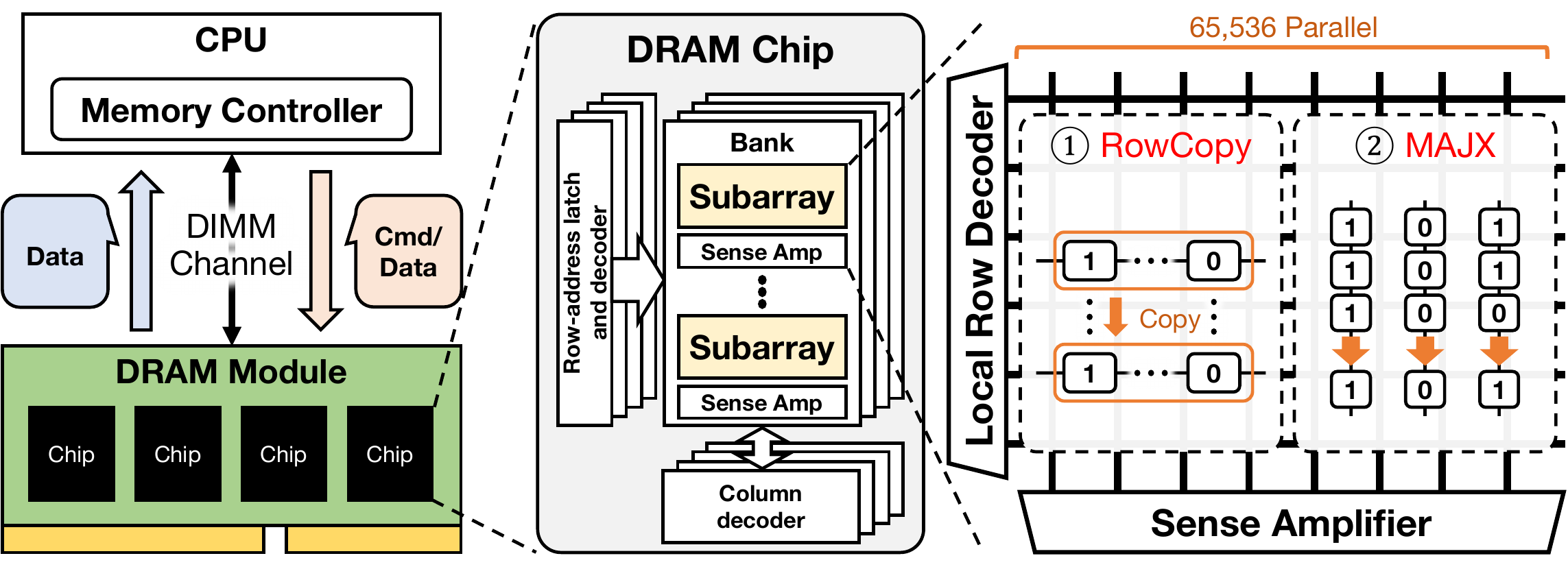}
    \caption{DRAM organization and PUD operations.}
    \label{fig:background1}
\end{figure}
\subsection{DRAM Fundamentals}
Dynamic Random Access Memory (DRAM) serves as the primary memory component in modern computing systems, providing high-density and cost-effective storage.
As shown in \Fig{fig:background1}, DRAM is organized in a hierarchical structure.
At the highest level, a DRAM system consists of multiple channels, each operating independently.
Each channel contains DRAM chips, which are further divided into banks.
Banks contain multiple \textit{subarrays}, each comprising a row decoder, sense amplifiers, and a grid of memory cells arranged in rows (256$\sim$1,024) and columns (65,536).
Each memory cell stores a single bit of data and is connected to a \textit{wordline} (row) and a \textit{bitline} (column), where wordlines activate rows of cells and bitlines transfer data between cells and sense amplifiers.

The memory controller orchestrates DRAM operations by issuing a sequence of commands to control data access.
The \ACT{} (Activation) command opens a specific row and copies its data to the row buffer, while the \PRE{} (Precharge) command closes the active row and prepares the bank for the next activation.
These commands follow specific timing constraints to maintain data integrity and reliability during normal DRAM operations~\cite{jedecDDR42012}.

\subsection{Processing-Using-DRAM}
Processing-Using-DRAM (PUD) leverages the inherent analog operational characteristics of DRAM to enable highly parallel bit-serial computations directly within memory arrays.
Prior research~\pudUnmodifiedCite{} has demonstrated that commercial off-the-shelf DRAM can achieve PUD functionality without hardware modifications by intentionally violating the timing parameters.

These studies have established two fundamental PUD operations: \RowCopy{} and \textit{majority-of-X} (\MAJ{X}) (\Fig{fig:background1}).
The \RowCopy{} operation facilitates data movement between different rows within a subarray by issuing a \PRE{} command followed immediately by an \ACT{} command before bitline precharging completes, enabling data transfer through the bitlines.
This operation affects all cells along a row simultaneously, making it approximately 100 times faster than processor-mediated data movement~\cite{seshadriRowCloneFastEnergyefficient2013}.
The \MAJ{X} operation performs a majority vote among $X$ cells sharing the same bitline that are activated simultaneously, implemented in commercial DRAM by issuing \ACT{}, \PRE{}, and \ACT{} commands in rapid succession without delays.
This allows concurrent activation of 2$\sim$32 rows. \MAJ{X} enables bit-serial computations that leverage the parallelism of subarrays with 65,536 columns, serving as the fundamental computational unit for PUD.

\begin{figure}[t]
    \centering
    \includegraphics[width=\columnwidth]{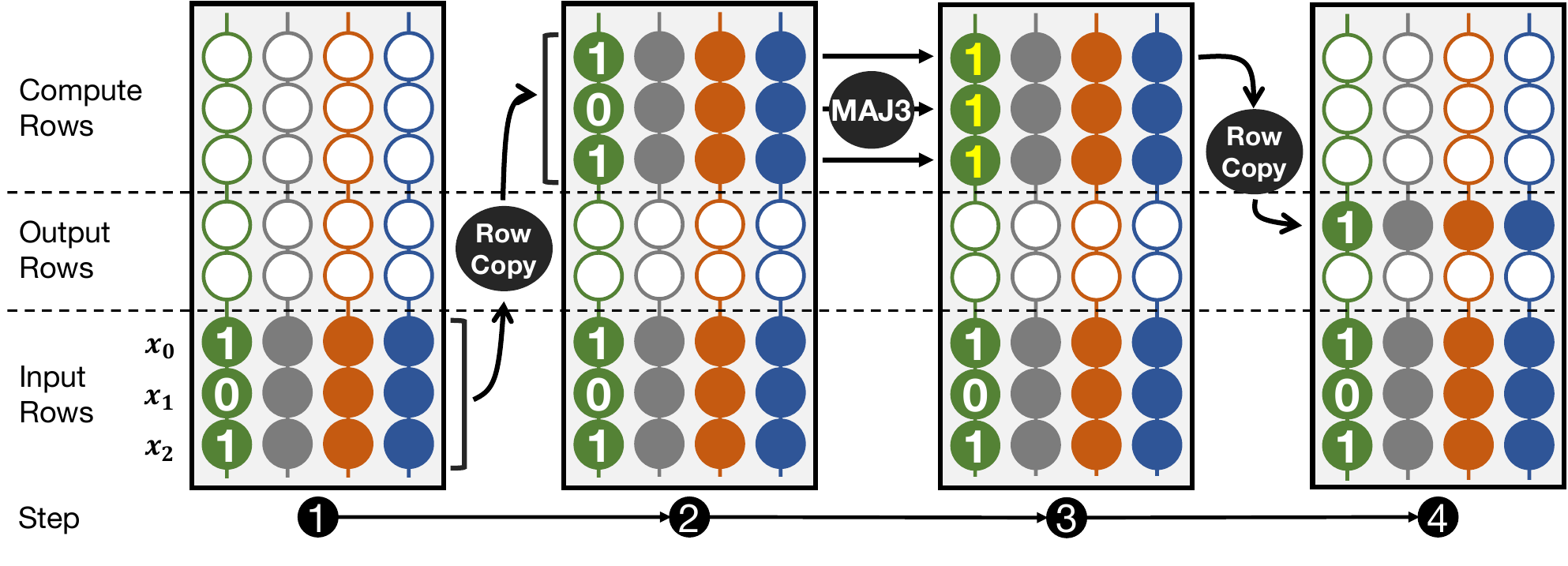}
    \caption{Data changes in a DRAM segment during a majority-of-3 operation.}
    \label{fig:background2}
\end{figure}
\Fig{fig:background2} illustrates a typical timeline of PUD operations.
The process consists of the following steps:
\circled{1} The input data is arranged on the same bitline to enable parallel computation across all 65,536 columns.
\circled{2} The input data is copied to computation rows to preserve the original inputs.
\circled{3} A \MAJ{3} operation is performed on the computation rows.
\circled{4} Finally, the result is copied to a designated destination row.
This sequence enables parallel bit-wise computation directly within the DRAM array while maintaining data integrity throughout the process.

\subsubsection{Full-Adder Calculation}\label{sec:background:full-adder}
Prior works~\cite{kuboBulkBitwiseAccumulation2024,aliInMemoryLowCostBitSerial2020} have proposed an implementation of full adder logic tailored for PUD, using \MAJ{X} and \NOT{} operations.
The implementation proceeds as follows:
\begin{align*}
    s_1 &= \text{MAJ}(x_0, x_1, x_2), \\
    s_0 &= \text{MAJ}(x_0, x_1, x_2, \bar{s_1}, \bar{s_1}),
\end{align*}
where $x_0, x_1, x_2$ are the three inputs to the full adder, and $s_1$ and $s_0$ are the carry-out and sum, respectively.
While some PUD works~\cite{seshadriAmbitInmemoryAccelerator2017} propose circuit-level modification methods for supporting \NOT{} operations, practical implementations in unmodified DRAM lack native \NOT{} operations\footnote{
    While prior work~\cite{yukselFunctionallyCompleteBooleanLogic2024} has proposed a \NOT{} implementation using unmodified DRAM, it is still limited to perform \NOT{} within a single subarray, preventing PUD from implementing Turing-complete logic.
}.
To address this limitation, we adopt a \textit{dual-track} approach, which is detailed in Section~\ref{sec:implementation}.

\subsubsection{Multiply-Accumulate Calculation}\label{sec:background:mac}
Prior works~\pudModifiedFrameCite{} have shown that PUD can achieve high computational throughput for multiply-accumulate (MAC) operations, which are fundamental building blocks for GeMV computations.
MAC operations are composed of multiplication and addition.
For multiplication in MAC operations, the calculation is decomposed into two steps: generating \textit{partial products} and then aggregating them.
Partial products are generated by multiplying each bit of one operand with each bit of the other operand, using bitwise \AND{} operations~\cite{gaoComputeDRAMInMemoryCompute2019}.
Below shows a $2$-bit by $2$-bit multiplication:
\[
    \begin{array}{r@{\quad}r@{\quad}r@{\quad}r@{\quad}r}
          &         &                &        a^{(1)} &        a^{(0)} \\
    \times&         &                &        w^{(1)} &        w^{(0)} \\
    \hline
          &         &                & a^{(1)}w^{(0)} & a^{(0)}w^{(0)} \\
    +     &         & a^{(1)}w^{(1)} & a^{(0)}w^{(1)} &                \\
    \hline
          & o^{(3)} &        o^{(2)} &        o^{(1)} &        o^{(0)}
    \end{array}
\]
PUD can implement this calculation by executing four bitwise \AND{} operations and two full-adder operation iteratively.
To implement a complete MAC operation, the multiplication result is further accumulated using additional full-adder stages.
These principles apply for signed arithmetic by properly handling two's complement bits~\cite{hennessy2011computer}.

PUD processes each MAC calculation entirely within a single column of the DRAM subarray.
When handling large-dimension MAC operations where the volume of data exceeds the available row count in a subarray (typically around 512 rows), computations must be distributed across multiple subarrays.
In this case, PUD systems compute partial sums independently within each subarray, which are then retrieved and aggregated by the processor to produce the final result.

 \section{Motivation}
The combination of PUD techniques and large low-bit GeMV operations in LLM inference presents an opportunity to transform on-device DRAM into an additional GeMV acceleration engine.
This approach leverages DRAM as a dual-purpose asset: it continues to serve as storage for model parameters while functioning as a computational resource for GeMV operations.
Moreover, the PUD techniques using unmodified DRAM eliminates the need for hardware modifications to existing memory devices.

\begin{figure}[h]
    \centering
    \includegraphics[width=\columnwidth]{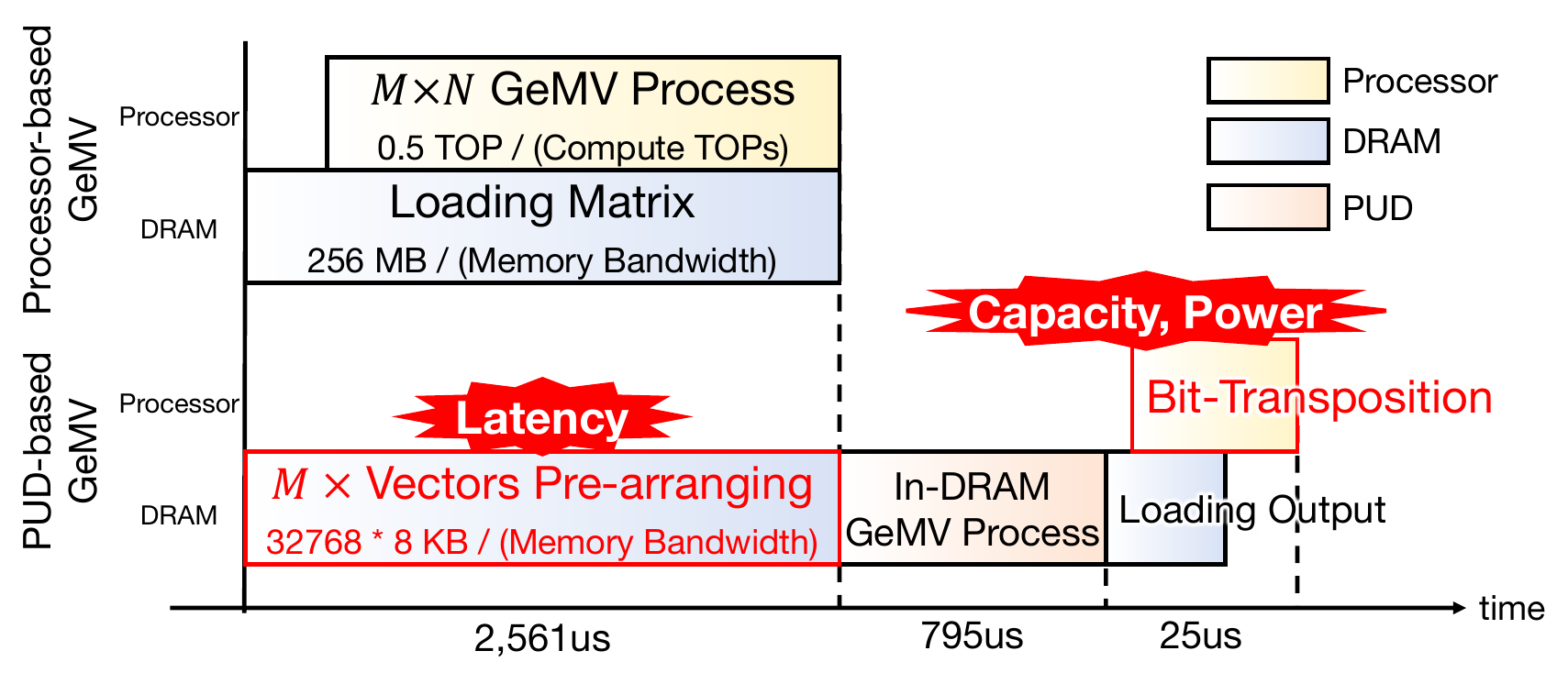}
    \caption{
        Latency profile comparison of GeMV operation between processor-based and PUD-based approaches.
        }
    \label{fig:motivation}
\end{figure}
However, applying PUD to GeMV operations in LLM inference introduces too much overhead \textit{before} and \textit{after} in-DRAM computation to be practically used in LLM inference.
These overheads stem from the fundermental challenge of PUD computation mechanism: \textit{the limitation of column-to-column data movement}, which is the inability of PUD to move data across different columns in a DRAM subarray.

\Fig{fig:motivation} compares the latency profile of a $32768 \times 8192$ $4$-bit GeMV calculation, whose dimension is typical in modern LLM inference, between processor-based and PUD approaches.
For clarity, throughout this paper, we use the notation $M \times N$ to refer to a GeMV operation involving an $M \times N$ matrix and an input vector of length $N$, resulting in an output vector of length $M$.
The figure shows results of our experimental evaluation using four DDR4-2400 DRAM modules.
The highlighted red portions illustrate the overheads in latency, capacity, and power consumption associated with conventional PUD approaches.

Before in-DRAM computation, conventional PUD requires pre-arranging the input vector in DRAM by $M$ times, which introduces substantial latency overheads for GeMV operations in LLM inference.
For an $M \times N$ GeMV operation, PUD needs to place $M$ combinations of the common input vector and different matrix vectors in DRAM to enable parallel MAC operations.
While matrix values can be pre-loaded in DRAM before inference, activation vector computation involves floating-point operations that must be performed on the processor during inference~\cite{hajinazarSIMDRAMFrameworkBitserial2021}.
Consequently, PUD must replicate and transfer the processor-generated activation vector $M$ times, incurring a matrix-size-proportional data transfer cost.

Conventional PUD generates bit-transposed outputs, limiting the PUD's parallelism and introducing capacity and power overheads.
Conventional PUD designs employ a \textit{vertical layout} that arranges all bits of a data element along the same bitline to overcome limitations in column-to-column data movement.
This approach assigns one column per computation, not fully utilizing the 65,536 parallelism of PUD in most LLM GeMV operations.
Besides, PUD operations overwrite data in all columns of an active row and prevent unused columns from storing other data, introducing capacity overhead.
Additionally, since DRAM hardware naturally accesses data row-wise, retrieving these vertically arranged outputs requires additional bit-transpose processing when transferring results to the processor, introducing power overheads.
 \begin{figure}[t]
    \centering
    \includegraphics[width=\columnwidth]{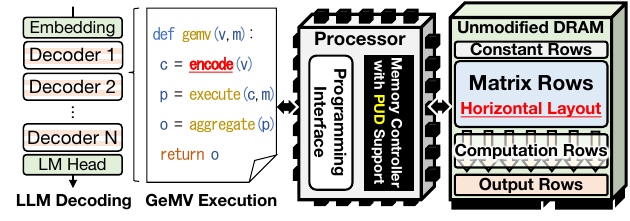}
    \caption{
        \MVDRAM{} system overview.
        }
    \label{fig:system}
\end{figure}
\section{\MVDRAM{} System}\label{sec:system}
This section outlines the \MVDRAM{} system, presenting its key components, novel techniques, execution flow, and memory organization.

\subsubsection*{\textbf{System Architecture}}
We design \MVDRAM{} on a standard computing platform comprising a processor and commercial off-the-shelf DRAM (\Fig{fig:system}).
The processor, which can be CPUs, GPUs, FPGAs, or another type of accelerator, manages the overall system and handles general-purpose computations.
The DRAM serves dual purposes: it stores the model state of the LLM and executes GeMV computations, leveraging PUD capabilities.
To leverage PUD techniques with real DRAM modules, \MVDRAM{} demands a memory controller that supports PUD-specific timing alongside conventional memory access.
Further discussion on \MVDRAM{}'s system integration and compatibility with other DRAM types is provided in Section~\ref{sec:discussion}.

\subsubsection*{\textbf{Key Techniques}}
\MVDRAM{} overcomes conventional PUD limitations through two novel techniques: \textit{on-the-fly vector encoding} and \textit{horizontal matrix layout}.
1) The on-the-fly vector encoding technique eliminates input pre-arranging overhead by directly encoding activation vector values into PUD operation sequences, leveraging the shared data patterns in GeMV operations to reduce data movement costs.
2) The horizontal matrix layout addresses capacity and power inefficiencies by organizing weight matrices row-wise and exploiting the mathematical linearity of matrix-vector multiplication, enabling enhanced parallelization and bit-transposition-free output aggregation.
These techniques are detailed in Section~\ref{sec:encoding} and Section~\ref{sec:layout}, respectively.

\subsubsection*{\textbf{Execution Flow}}
\MVDRAM{} executes GeMV operations through a four-step process:
\circled{1} \MVDRAM{} stores weight matrices in DRAM prior to computation.
These pre-loaded matrices remain in DRAM throughout the inference process.
\circled{2} When a GeMV operation is triggered, \MVDRAM{} encodes the processor-computed activation vector values directly into DRAM command sequences.
\circled{3} The encoded DRAM command sequences are issued to the DRAM modules through the PUD interface, executing partial GeMV computations in-DRAM.
\circled{4} After the in-DRAM computation completes, \MVDRAM{} aggregates the partial sums generated within DRAM to produce the final GeMV output vector.

\subsubsection*{\textbf{Memory Organization}}
To implement GeMV operations, \MVDRAM{} reserves four dedicated regions in the DRAM subarray:
1) The \textit{constant rows} comprise two rows reserved for PUD-based logical operations, one holding all zeros and the other all ones across all columns.
2) The \textit{matrix rows} are designated to hold matrix values.
They are referenced as inputs for PUD operations through \RowCopy{}.
3) The \textit{computation rows} store intermediate results of GeMV operations within DRAM.
Its values are repeatedly overwritten during computation.
4) The \textit{output rows} holds the final results of the GeMV operations computed in DRAM.
Upon completion, the processor loads data from this region to obtain the final output.
 \section{On-the-fly Vector Encoding}\label{sec:encoding}
\subsection{Challenge}
Conventional PUD approaches require pre-arranging input data before in-DRAM computation.
PUD operations execute through DRAM commands that specify addresses, with each command affecting all data stored in the specified rows.
When processor-computed values need to be incorporated into PUD operations, these values must first be transferred from the processor to DRAM, introducing significant data movement overhead.

While implementing the entire LLM inference exclusively within DRAM could eliminate the data transfer overhead, PUD's data movement constraints and bit-serial computation model make it impractical because some operations in LLM inference, such as softmax and layer normalization, require access to all input data and floating-point arithmetic. 
Consequently, in LLM inference, computation is distributed between DRAM and the processor.

\begin{figure}[t]
    \centering
    \includegraphics[width=\columnwidth]{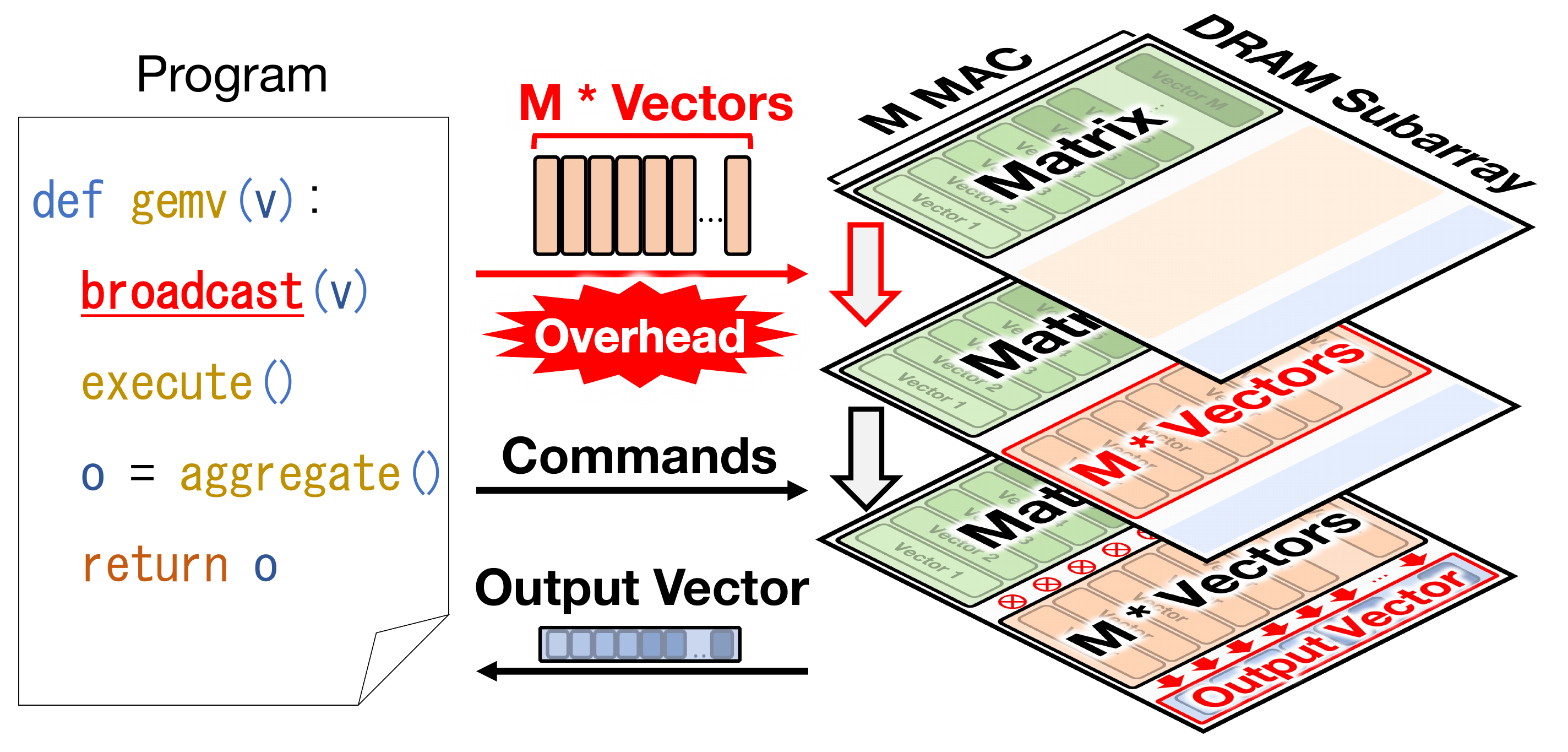}
    \caption{GeMV execution using conventional PUD.}
    \label{fig:encoding1}
\end{figure}
However, arranging the processor-computed input vector for GeMV operations introduces data transfer costs proportional to the matrix size.
Due to the limited column-to-column data movement capability of PUD, for an $M \times N$ GeMV operation, PUD needs to place $M$ combinations of the common vector with different matrix vectors in DRAM to enable parallel MAC operations (\Fig{fig:encoding1}).
While matrix values can be pre-loaded before GeMV execution, processor-computed input vector must be transferred to DRAM during runtime.
This necessitates replicating and transferring the vector $M$ times, resulting in data transfer costs proportional to the matrix size and diminishing the benefits of in-DRAM computation.

\subsection{Insight}
GeMV operations share the same input vector across all parallel MAC operations.
The MAC operations in an $M \times N$ GeMV, which comprise $M$ computations between $M$ unique matrix vectors and a common input vector, rely on the same input vector for all computations. 
This shared data pattern presents an opportunity to optimize the PUD command sequences in two ways:
(1) The fixed pattern can be directly encoded into the PUD operation sequence, eliminating the need to transfer the vector from the processor to DRAM.
(2) Our method can reduce computational complexity by skipping operations when activation bits are 0, further optimizing performance for the existing diverse sparse activation patterns~\cite{gaoSeerAttentionLearningIntrinsic2025,wangQSparseAllLarge2024,liuTrainingFreeActivationSparsity2025}.

\subsection{Method}
To eliminate the pre-arranging cost of the input vector, we propose \textit{on-the-fly vector encoding}, a technique that embeds the input vector's values directly into PUD operations.
Unlike conventional PUD, which issues a fixed command sequence independent of the input data, this approach dynamically generates a command sequence based on the input vector.
By issuing this command sequence, \MVDRAM{} can execute the required GeMV computations in-DRAM without replicating and transferring the input vectors to DRAM.

The proposed on-the-fly vector encoding specifically targets the partial product calculations within MAC operations.
As outlined in Section~\ref{sec:background:mac}, PUD implements MAC operations through a combination of partial product computations and full adder operations.
Each partial product is computed as the \AND{} operation between a bit of the weight and a bit of the input, with full adders iteratively applied to aggregate these products.
Our encoding replaces these partial product computations with \RowCopy{} operations dynamically encoded based on the input vector's bit pattern.

\begin{figure}[t]
    \centering
    \begin{subfigure}[b]{0.3\columnwidth}
        \centering
        \includegraphics[width=0.8\columnwidth]{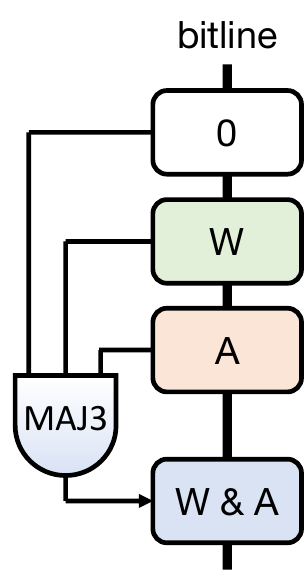}
        \caption{Prior PUD.}
        \label{fig:encoding2_1}
    \end{subfigure}
    \begin{subfigure}[b]{0.6\columnwidth}
        \centering
        \begin{subfigure}[b]{0.4\columnwidth}
            \centering
            \includegraphics[width=0.9\columnwidth]{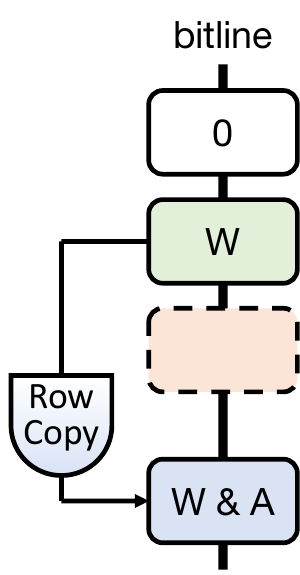}
            \caption*{$a = 1$}
        \end{subfigure}
        \begin{subfigure}[b]{0.4\columnwidth}
            \centering
            \includegraphics[width=0.9\columnwidth]{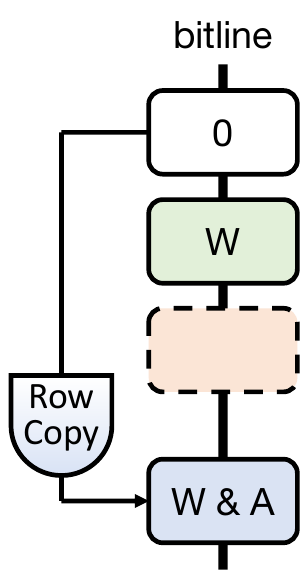}
            \caption*{$a = 0$}
        \end{subfigure}
        \caption{On-the-fly vector encoding.}
        \label{fig:encoding2_2}
    \end{subfigure}
    \caption{Partial product generation using conventional PUD and on-the-fly vector encoding.}
    \label{fig:encoding2}
\end{figure}
\Fig{fig:encoding2} compares the partial product implementation in conventional PUD and on-the-fly vector encoding.
Typical PUD realizes the partial product using a \MAJ{3} operation that takes two input rows (containing the matrix bit ($w$) and vector bit ($a$)) and a constant zero row as inputs, computing the \AND{} product within a column~\cite{seshadriAmbitInmemoryAccelerator2017,gaoComputeDRAMInMemoryCompute2019} (\Fig{fig:encoding2_1}).
In contrast, our approach leverages selective \RowCopy{} operations dictated by the vector bit value (\Fig{fig:encoding2_2}).
When the vector bit $a = 1$, the row storing the matrix bit is copied to the target row, resulting in an output of $o = 1$ only if the matrix bit $w = 1$.
When $a = 0$, a constant zero row is copied to the target row, producing an output of $o = 0$ regardless of the matrix bit's value.
By dynamically adjusting the source row in these \RowCopy{} commands, we implement the partial product without storing the vector values in DRAM.

To implement on-the-fly vector encoding, the encoding method operates by scanning each bit of the vector elements and replacing the corresponding addresses in a pre-prepared command template. 
Since the encoding only modifies the reference addresses of the \RowCopy{} operations, \MVDRAM{} can prepare the overall command sequence structure for executing MAC operations in advance. 
This approach maintains a computational complexity of $O(n)$ relative to the vector length.
Once encoded, these PUD command sequences are sent to the DRAM through the processor's interface.

\subsection{Bit Sparsity Optimization}
Our on-the-fly vector encoding technique further leverages vector bit sparsity to reduce the number of DRAM operations.
When a vector bit is 0, rather than explicitly copying data from a constant zero row, \MVDRAM{} can simply skip the operation entirely.
This optimization not only reduces the number of \RowCopy{} commands but also decreases the subsequent full adder execution count, leading to significant latency improvements for sparse vectors.

To implement this optimization without increasing runtime encoding overhead, we prepare multiple command sequence templates based on potential bit count patterns.
During execution, \MVDRAM{} selects the appropriate template based on the total number of set bits in the input vector, then replaces only the necessary addresses.

\subsection{Overhead}
The dynamic encoding overhead is negligible as long as command generation throughput exceeds DRAM's command processing rate.
While a DDR4-2400 DRAM module processes approximately one command every $1.5\si{\ns}$, our preliminary evaluation shows that even a single-threaded processor implementation can generate commands faster.
By overlapping command generation with execution, \MVDRAM{} effectively masks any overhead from dynamic command generation. \begin{figure}[t]
    \centering
    \begin{subfigure}[b]{0.48\columnwidth}
        \centering
        \includegraphics[width=0.8\columnwidth]{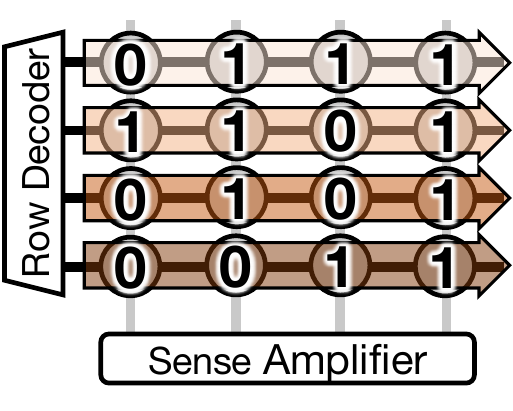}
        \caption{Horizontal data access.}
        \label{fig:layout1_1}
    \end{subfigure}
    \begin{subfigure}[b]{0.48\columnwidth}
        \centering
        \includegraphics[width=0.8\columnwidth]{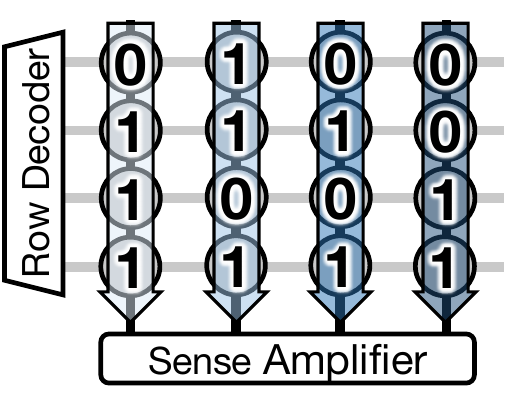}
        \caption{PUD's vertical layout.}
        \label{fig:layout1_2}
    \end{subfigure}
    \caption{
        Horizontal data access and PUD's vertical layout where four 4-bit numbers of $(0111_2, 1101_2, 0101_2, 0011_2)$ are stored.
}
    \label{fig:layout}
\end{figure}
\section{Horizontal Matrix Layout}\label{sec:layout}
\subsection{Challenge}
Due to the limitations of data movement between columns, conventional PUD approaches employ a \textit{vertical layout} strategy~\cite{hajinazarSIMDRAMFrameworkBitserial2021}.
Unlike the conventional horizontal data layout in DRAM where all bits of a data element are arranged in the same row (\Fig{fig:layout1_1}), PUD transposes these bits and places them in the same column (Figure~\ref{fig:layout1_2}).
This vertical arrangement enables PUD to perform bit shift operations via \RowCopy{}, facilitating multi-bit computations.

However, this vertical layout, which assigns one column per computation, degrades the PUD's parallelism in GeMV calculations.
For most GeMV operations in modern LLMs, the computational parallelism required is smaller than that available in PUD, which can process up to 65,536 columns simultaneously.
Besides, since PUD operations affect all data in the same row, any data in unused columns is destroyed during computation.
Consequently, systems with PUD cannot effectively utilize these columns for storing other data, leading to capacity overhead.

Additionally, since DRAM is designed for row-wise data access, the vertical layout introduces additional power overhead due to the processor's bit-transposition.
DRAM accesses data by row through the subarray's row decoder and sense amplifiers.
Consequently, outputs generated in a vertical layout require additional bit-transpose processing when transferred to the processor.
The processor must buffer the row-wise read data and apply bit-transposition processing to obtain the final PUD outputs.
SIMDRAM~\cite{hajinazarSIMDRAMFrameworkBitserial2021} proposes to add a bit transpose unit on-chip, reporting up to 91\% additional latency overhead.

\subsection{Bit-Decomposition of MAC}
To address the overheads introduced by vertical layout, we leverage the mathematical linearity inherent in MAC operations.
This linearity allows us to decompose MAC operations into independent computations that can be performed separately and then combined to obtain the original result.
Implementing general non-linear computations in PUD typically requires arranging input data vertically (placing all bits in a single column) due to limitations in data movement between columns.
In contrast, the linearity of MAC operations presents an opportunity to distribute computations across different columns and interpret results row-wise, offering enhanced parallelism and efficient reconstruction of the final output without bit-transposition.

\begin{figure}[h]
    \centering
    \includegraphics[width=\columnwidth]{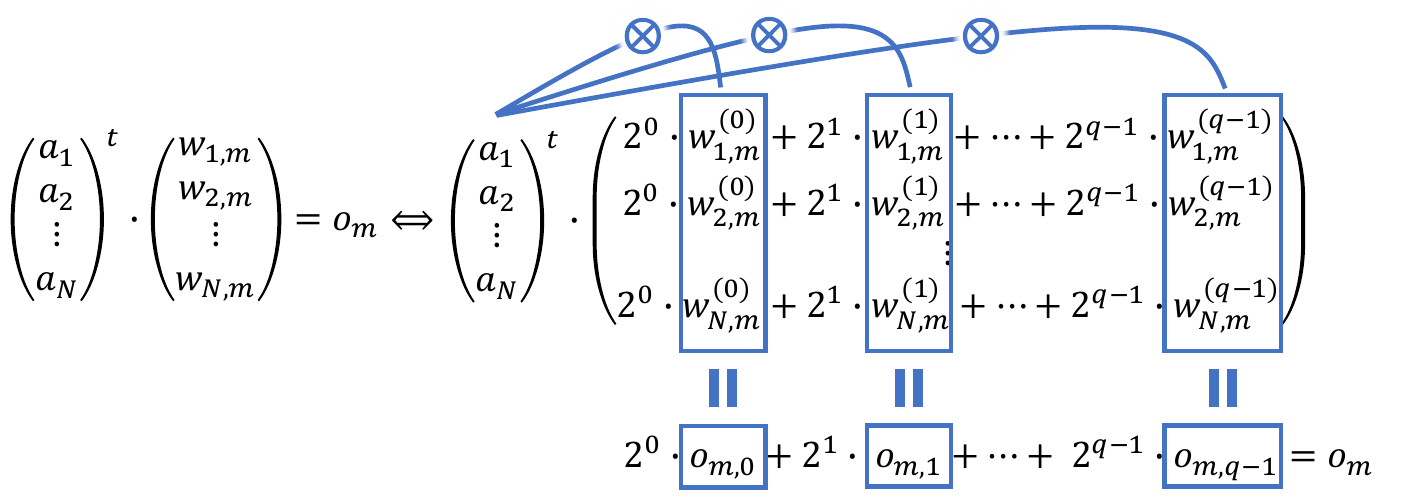}
    \caption{Bit-decomposition of MAC operation with respect to the matrix bits.}
    \label{fig:layout2}
\end{figure}
\MVDRAM{} applies bit-decomposition to MAC operations with respect to the matrix bits (\Fig{fig:layout2}).
In an $M \times N$ GeMV, $M$ MAC operations between the input vector and each vector of the matrix are performed, each computing the accumulated sum of $N$ partial sums.
For the $m$-th MAC result $o_m$, this is expressed as: $o_m = \sum_{i=1}^N a_i w_{m,i}$.
For matrix element of $q$-bit precision, which we denote as $w_{m,n} = \sum_{i=0}^{q-1} 2^i w_{m,n}^{(i)}$, our bit-decomposition method splits each MAC into $q$ partial sums ($o_{m,0}, o_{m,1}, \ldots, o_{m,q-1}$).
Each partial sum $o_{m,i}$ represents the MAC between the $N$ activations and the $i$-th bit of the matrix element.
This is expressed as:
\begin{align*}
    o_{m,i} &= \sum_{j=1}^N a_j w_{m,j}^{(i)}.
\end{align*}
These partial sums are weighted by their corresponding bit positions (i.e., $2^i$) and summed to reconstruct the full MAC result:
\begin{align*}
    \sum_{i=0}^{q-1} 2^i o_{m,i}
        &= \sum_{i=0}^{q-1} 2^i \sum_{j=1}^N a_j w_{m,j}^{(i)} \\
        &= \sum_{j=1}^N a_j \left( \sum_{i=0}^{q-1} 2^i w_{m,j}^{(i)} \right) \\
        &= \sum_{j=1}^N a_j w_{m,j} \\
        &= o_m.
\end{align*}
The point is that this weighted accumulation naturally aligns with DRAM's horizontal access pattern for multi-bit data retrieval.
For example, computing the value $(1011)_2 = 11$ involves accumulating each bit with its corresponding matrix element $(2^0 + 2^1 + 2^3)$, which is precisely how processors naturally interpret multi-bit values stored horizontally in DRAM.

\subsection{Method}
\begin{figure}[t]
    \centering
    \begin{subfigure}[b]{\columnwidth}
        \centering
        \includegraphics[width=\columnwidth]{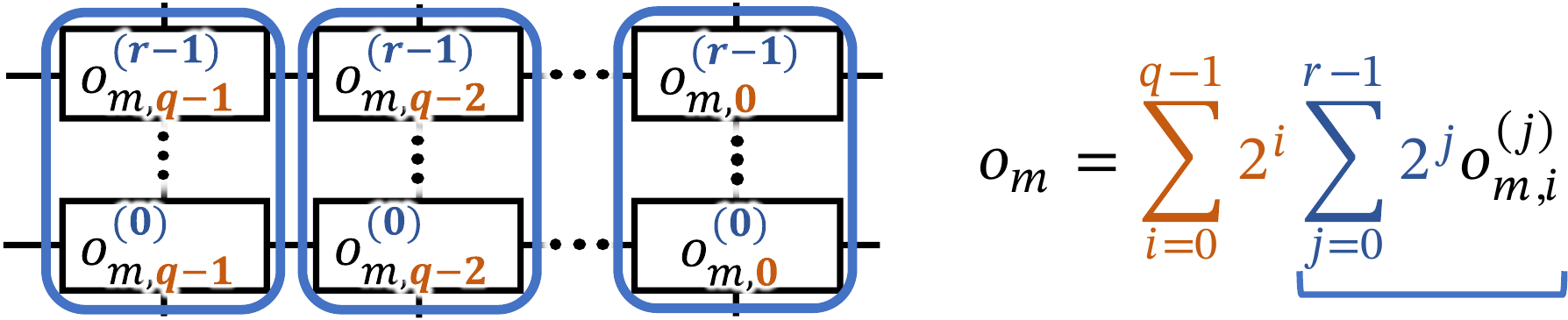}
        \caption{Column-wise interpretation of the output $o_{m}$.}
        \label{fig:layout3_1}
    \end{subfigure}
    \begin{subfigure}[b]{\columnwidth}
        \centering
        \includegraphics[width=\columnwidth]{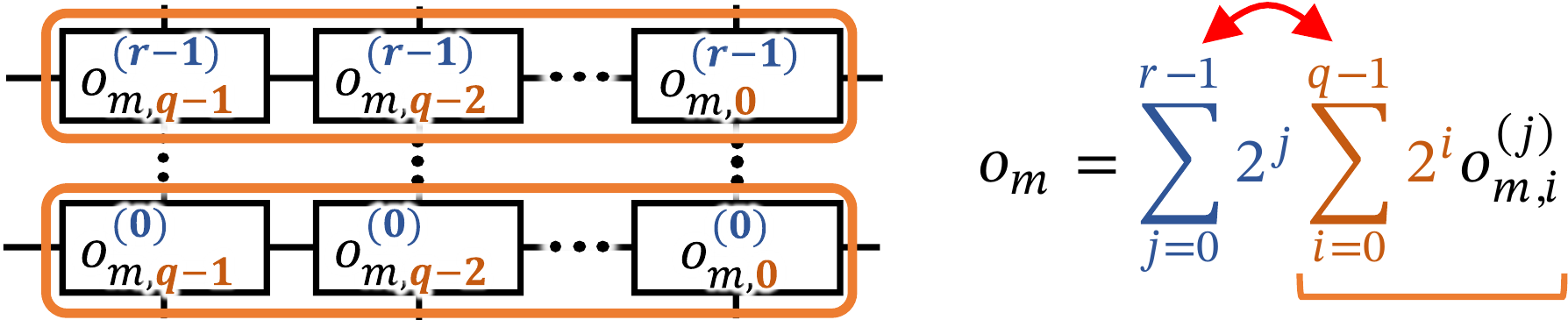}
        \caption{Row-wise interpretation of the output $o_{m}$.}
        \label{fig:layout3_2}
    \end{subfigure}
    \caption{Column-wise and row-wise interpretations of $o_{m}$.}
    \label{fig:layout3}
\end{figure}
\begin{figure}[t]
    \centering
    \includegraphics[width=\columnwidth]{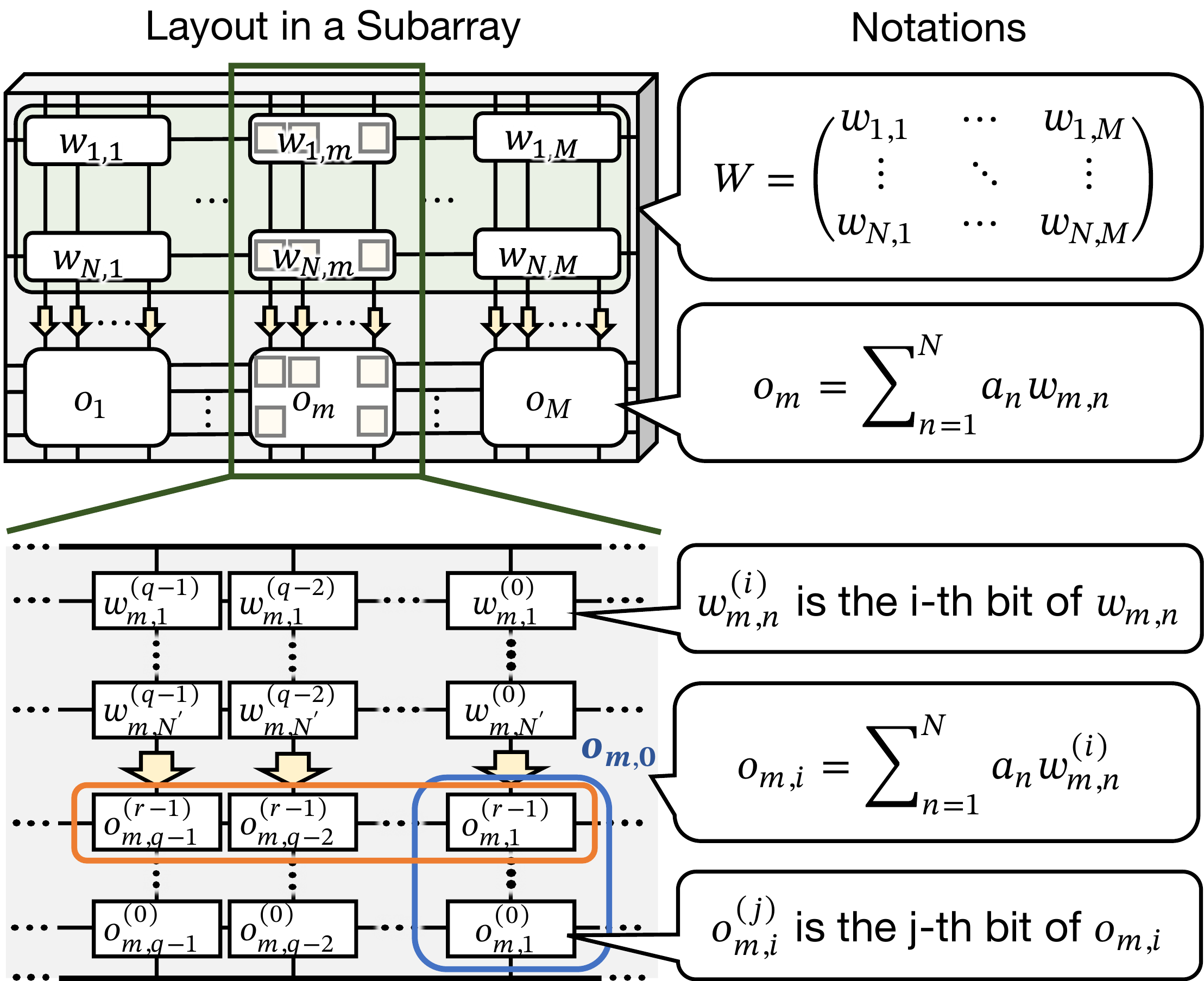}
    \caption{Horizontal matrix layout.}
    \label{fig:layout4}
\end{figure}
Based on the bit-decomposition of MAC operations, we propose a \textit{horizontal matrix layout} that enhances the PUD's column utilization and aligns with DRAM's natural access patterns.
The key insight is that the weighted accumulation of these independently computed partial sums aligns with how multi-bit data is naturally retrieved through DRAM's horizontal access mechanism.
\Fig{fig:layout3} illustrates the difference between conventional column-wise access and our row-wise interpretation of outputs.
\Fig{fig:layout3_1} shows how the output of each bit-decomposed MAC operation is stored.
When the output value has $r$ bits, $r$ rows are consumed to store the bit-decomposed MAC values.
This conventional column-wise access requires bit transposition to reconstruct the final output.
In contrast, \Fig{fig:layout3_2} demonstrates how our approach reinterprets these outputs horizontally.
By accessing $q$-bit values horizontally and performing weighted accumulation with respect to the output bits, \MVDRAM{} can efficiently obtain the final MAC outputs without explicit bit-transposition operations.

\Fig{fig:layout4} illustrates the \MVDRAM{}'s mapping of an $M \times N$ dimensional $q$-bit matrix and $M$ dimensional $r$-bit output vector within a subarray.
In the row direction, we store all bits related to different values along the $M$ dimension of the matrix.
This arrangement utilizes $q M$ columns, enabling computational parallelism with $q M$ simultaneous operations.
In the column direction, we organize different values along the $N$ dimension of the matrix, with corresponding bit positions sharing the same bitline.
The output values are expanded and arranged as $q \times r$ bits.
By accessing these outputs row-wise and performing shift accumulation across $r$ rows, processors can efficiently compute the final output values $o_m$ without requiring explicit bit-transposition operations.

\subsection{Advantages}
Our horizontal matrix layout offers two key benefits over conventional PUD implementations:
(1) By organizing weight bits horizontally, we enable $q M$ simultaneous operations instead of just $M$, better utilizing PUD's 65,536-column parallelism.
(2) Outputs align with DRAM's natural row-wise access, eliminating bit-transposition operations and allowing direct interpretation as multi-bit values.
 \section{\MVDRAM{} Implementation}\label{sec:implementation}
\begin{figure}[t]
    \centering
    \includegraphics[width=0.8\columnwidth]{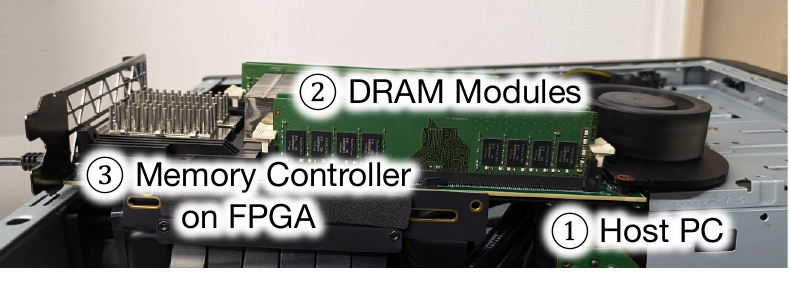}
    \caption{\MVDRAM{} system setup.}
    \label{fig:setup}
\end{figure}
\subsubsection*{\textbf{Setup}}
\Fig{fig:setup} illustrates the experimental setup of \MVDRAM{}.
This configuration consists of \circled{1} a host PC, \circled{2} DRAM modules, and \circled{3} a memory controller implemented on an FPGA.
Instead of accessing memory through the processor's native memory controller, \MVDRAM{} bypasses it to directly control the DRAM modules via the FPGA-based memory controller.
This approach allows us to implement the precise timing parameters required for PUD operations on commercial DRAM modules.
We use DRAM Bender~\cite{olgunDRAMBenderExtensible2023}, an open-source memory controller framework, implemented on a Xilinx Alveo U200 FPGA.
We use SK Hynix DDR4-2400 memory modules\footnote{
    We used four DRAM modules HMA851U6CJR6N-UHN0.
    Through the characterization of 16 different SK Hynix DRAM modules, we identified this specific part number as the most reliable one that supports both strict \RowCopy{} and \MAJ{X} operations (up to \MAJ{15}).
} for all \MVDRAM{} experiments to support reliable PUD operations.

\subsubsection*{\textbf{Matrix Partitioning}}
To handle large-dimension GeMV operations, we partition the matrix and distribute them across multiple DRAM modules and subarrays.
Within each subarray, we limit the maximum $N$ dimension to $128$ and partition it across multiple subarrays so that we can accommodate all rows required for computation within a single subarray.
Additionally, when the product of weight bit-width $q$ and $M$ exceeds the available column count, we distribute operations across additional subarrays.

\subsubsection*{\textbf{Dual-Track Approach}}
To address the lack of native \NOT{} operations in unmodified DRAM, we employ a \textit{dual-track} approach~\cite{kuboBulkBitwiseAccumulation2024} that maintains both original and complementary values throughout computation.
For the full adder implementation, we prepare both inputs and their complements, compute the carry-out ($s_1$) and sum bit ($s_0$) using \MAJ{3} and \MAJ{5} operations respectively, along with their complements.
This strategy enables complete logical operations using only \RowCopy{} and \MAJ{X} primitives available in unmodified DRAM, though at the cost of additional row usage which we evaluate in Section~\ref{sec:evaluation:results}.

\begin{table}[t]
    \centering
    \caption{\# of reliable columns.}
    \label{tab:reliable}
    \begin{tabularx}{\linewidth}{p{0.3\linewidth}c}
    \toprule
    \textbf{DRAM} & \textbf{Min. \# of reliable columns} \\
    \midrule
    Module \#1   & 61,727 (max. 62,826) / 65,536   \\
    Module \#2   & 62,300 (max. 62,483) / 65,536   \\
    Module \#3   & 54,365 (max. 62,329) / 65,536   \\
    Module \#4   & 54,712 (max. 62,925) / 65,536   \\
    \bottomrule
    \end{tabularx}
\end{table}
\subsubsection*{\textbf{Reliable MAJX}}
PUD's MAJX operations inherently contain errors in some columns of commercial DRAM modules~\cite{gaoComputeDRAMInMemoryCompute2019,yukselSimultaneousManyRowActivation2024}.
To address this reliability challenge, \MVDRAM{} employs \texttt{Frac} operations~\cite{gaoFracDRAMFractionalValues2022} and calibration techniques~\cite{yu65nm8TSRAM2022} to increase the number of reliable columns, which achieves error-free computation.
The number of reliable columns is shown in Table~\ref{tab:reliable}.
For our implementation, we use only consecutive sequences of $q$ reliable columns when performing $q$-bit GeMV operations to ensure error-free computation.
While this selection introduces a slight data transfer overhead for unused columns, the impact on aggregation latency is minimal (\Fig{fig:motivation}).
 \section{Evaluation}
\subsection{Methodology}\label{sec:methodology}

\begin{table}[t]
    \centering
    \caption{System platforms.}
    \label{tab:platforms}
\begin{tabularx}{\linewidth}{p{0.23\linewidth}cc}
    \toprule
    \textbf{Platform} & \textbf{DRAM} & \textbf{Processor} \\
    \midrule
    {\small Baseline (CPU)}   & {\small DDR4-2400 (77 GB/s)} & {\small Intel Core i7-9700K}   \\
    {\small Baseline (GPU)} & {\small LPDDR5 (68 GB/s)}    & {\small NVIDIA Jetson Orin Nano}  \\
    {\small \MVDRAM{}}    & {\small DDR4-2400 (77 GB/s)} & {\small Intel Core i7-9700K} \\
    \bottomrule
    \end{tabularx}
\end{table}
\subsubsection*{\textbf{Platforms}}
We evaluate \MVDRAM{} against two baseline platforms to demonstrate its performance advantages over conventional processor-based implementations.
Table~\ref{tab:platforms} summarizes the specifications of the three platforms used in our evaluation.
For fair comparison, both the CPU and \MVDRAM{} implementations use the DDR4-2400 memory modules and Intel Core i7-9700K processor.

\subsubsection*{\textbf{Benchmarks}}
Our evaluation focuses on two benchmarks: GeMV operations and LLM inference.
For GeMV benchmarks, we test matrix dimensions used in modern LLMs, with weight precision from $2$-bit to $8$-bit.
All performance measurements represent averages across 1,000 iterations, using different input values between iterations to prevent cache optimization effects.
We use input vectors with 50\% bit sparsity, which represents a typical distribution in practical LLM workloads~\cite{wangQSparseAllLarge2024,liuTrainingFreeActivationSparsity2025}.
We use the \texttt{ggml}~\cite{ggml} library to implement GeMV operations with quantized weights on CPU and GPU platforms.

For end-to-end LLM inference, we benchmark token generation throughput across four representative models: Llama2-7B~\cite{touvronLlama2Open2023}, Llama2-13B~\cite{touvronLlama2Open2023}, Llama3-8B~\cite{llama3.1405B}, and Phi-4~\cite{abdinPhi4TechnicalReport2024}.
Our end-to-end LLM implementation builds on \texttt{llama.cpp}~\cite{llama_cpp}, where we replace \texttt{mulmat\_op} operations with our \MVDRAM{} implementation.
We measure the time required to generate $256$ tokens and compute the average throughput across $10$ repeated runs.

\subsection{Results}\label{sec:evaluation:results}
\subsubsection*{\textbf{GeMV}}
\begin{figure}[h]
    \centering
    \includegraphics[width=\columnwidth]{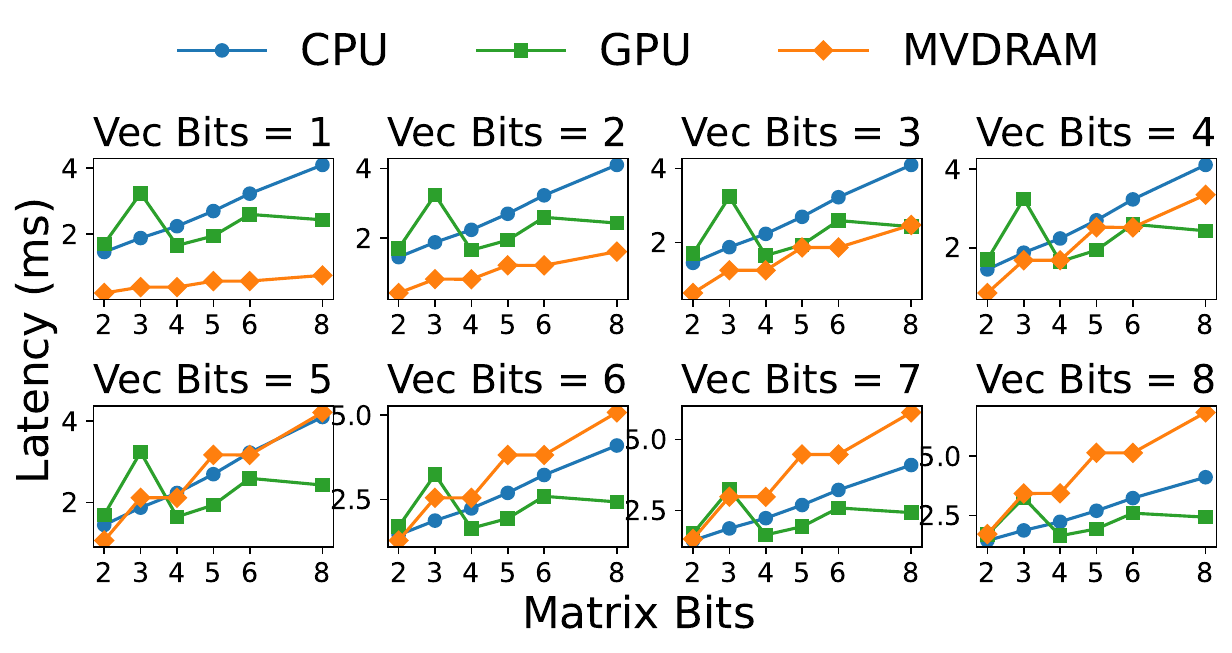}
    \caption{GeMV latency on different bit-widths.}
    \label{fig:evaluation1_1}
\end{figure}
\Fig{fig:evaluation1_1} shows the latency of a $32000 \times 4096$ GeMV operation, which is used in llama2-7B~\cite{touvronLlama2Open2023}, across different bit precisions.
We can observe that \MVDRAM{} achieves up to \gemvLatencyOverCpu{} and \gemvLatencyOverGpu{} speedup over CPU and GPU, respectively, when $1$-bit vector and $2$-bit matrix.
For this configuration, CPU and GPU implementations complete the operation in 1.44 $\si{ms}$ and 1.70 $\si{ms}$ respectively.
In contrast, MVDRAM completes the in-DRAM computation in just 0.14 $\si{ms}$ and result aggregation (for 32000$\times$384 bits) in 0.05 $\si{ms}$, totaling only 0.19 $\si{ms}$ for the entire operation.
This performance improvement stems from MVDRAM's highly parallel bit operations within DRAM, outperforming the matrix loading time required by conventional processors especially for large, low-bit matrix operations.

\begin{figure}[h]
    \centering
    \includegraphics[width=\columnwidth]{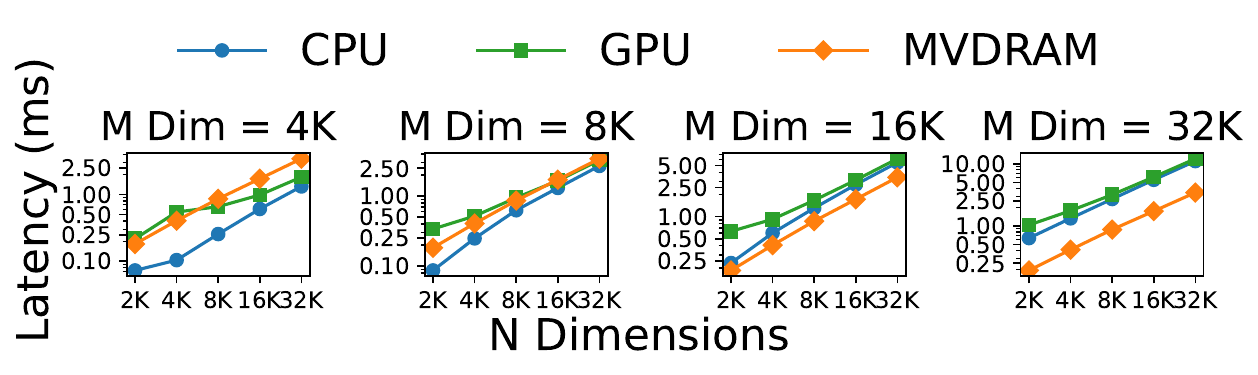}
    \caption{GeMV latency on different sizes in log scale.}
    \label{fig:evaluation1_2}
\end{figure}
\Fig{fig:evaluation1_2} illustrates the GeMV latency across different matrix dimensions ranged from 2,048 to 32,768 using $2$-bit precision.
At the largest dimensions ($M=32768, N=32768$), we can observe that \MVDRAM{} achieves $3.38\times$ and $3.74\times$ speedup compared to CPU and GPU, respectively.
This increasing performance advantage with larger matrix sizes demonstrates how \MVDRAM{} effectively leverages its high parallelism across thousands of DRAM columns.

\begin{figure}[h]
    \centering
    \includegraphics[width=\columnwidth]{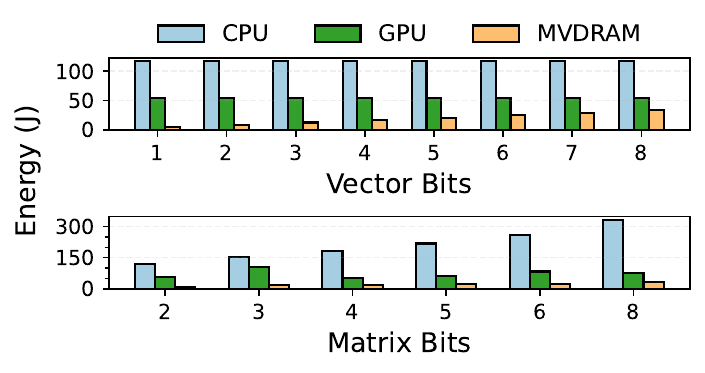}
    \caption{GeMV energy consumption.}
    \label{fig:evaluation1_3}
\end{figure}
To measure power consumption, we employed \texttt{Intel RAPL}~\cite{hahnelMeasuringEnergyConsumption2012} for CPU measurements, \texttt{tegrastats}~\cite{tegrastats} for GPU power, and CACTI~\cite{balasubramonianCACTI7New2017} for \MVDRAM{} power estimation.
For fair comparison, we normalized the GPU's energy consumption by accounting for the difference between LPDDR5 and DDR4 technologies, replacing the amount of LPDDR5 energy consumption with equivalent DDR4 energy.
\Fig{fig:evaluation1_3} presents the energy consumption of a $32000 \times 4096$ GeMV operation, across different bit precisions, with one operand fixed at $2$-bit width.
\MVDRAM{} demonstrates consistently superior energy efficiency across all configurations tested.
When vector bit-width is 1, it can be seen that \MVDRAM{} achieves energy efficiency improvements of \gemvEnergyOverCpu{} and \gemvEnergyOverGpu{} compared to CPU and GPU, respectively.
This energy advantage comes from reducing the data movement between memory and computation units, which is a significant source of energy consumption in conventional systems.

\begin{figure}[h]
    \centering
    \includegraphics[width=\columnwidth]{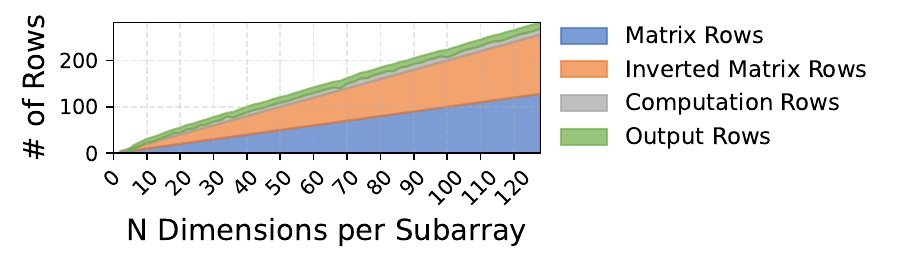}
    \caption{Capacity overhead.}
    \label{fig:evaluation1_4}
\end{figure}
\Fig{fig:evaluation1_4} illustrates the row utilization breakdown per subarray for $4$-bit GeMV operations across various $N$ dimensions.
The matrix rows and inverted matrix rows represent the storage requirements for weight data, while computation rows store intermediate results, and output rows contain the final results.
As shown in the figure, the overhead associated with computation and output rows remains consistently minimal compared to the matrix storage requirements, regardless of the dimension size.
This demonstrates that the capacity overhead introduced by MVDRAM's in-DRAM computation is negligible relative to the storage required for the matrix data.

\subsubsection*{\textbf{End-to-End LLM Inference}}
\begin{figure}[h]
    \centering
    \includegraphics[width=\columnwidth]{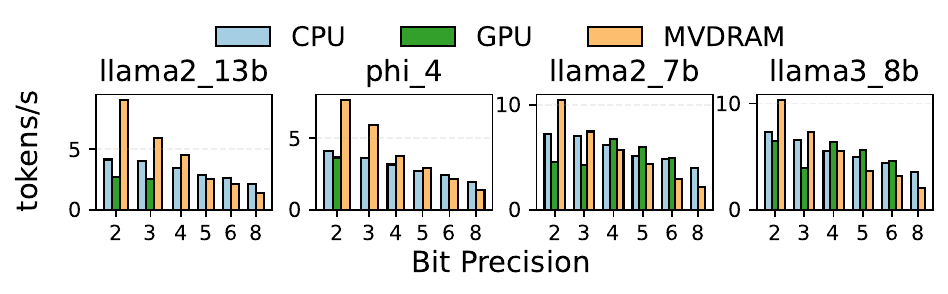}
    \caption{Token throughput. GPU results for configurations exceeding the memory limit (8GB) are omitted.}
    \label{fig:evaluation2_1}
\end{figure}
\Fig{fig:evaluation2_1} illustrates the token throughput of low-bit LLM inference across different bit precisions.
For $2$-bit Llama2-13B, we can observe that \MVDRAM{} achieves \llmTwoThroughputOverCpu{} and \llmTwoThroughputOverGpu{} higher throughput compared to CPU and GPU implementations, respectively.
With $4$-bit Llama2-13B, which are currently common in production environments, \MVDRAM{} still maintains a \llmFourThroughputOverCpu{} throughput advantage over CPU implementations.
These results highlight \MVDRAM{}'s effectiveness for current $4$-bit quantized models while suggesting even greater performance benefits for emerging $2$-bit quantization techniques that are expected to become more prevalent in future LLM deployments.

\begin{figure}[h]
    \centering
    \includegraphics[width=\columnwidth]{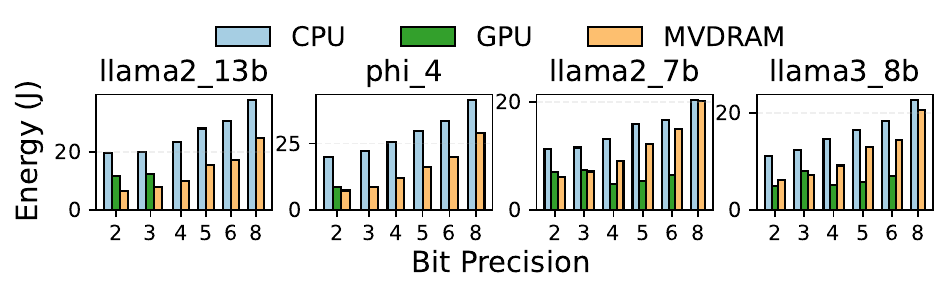}
    \caption{Energy consumption per token. GPU results for configurations exceeding the memory limit (8GB) are omitted.}
    \label{fig:evaluation2_2}
\end{figure}
\Fig{fig:evaluation2_2} presents the energy consumption per token for LLM inference across bit precisions.
For 2-bit Llama2-13B, \MVDRAM{} demonstrates energy efficiency improvements of \llmTwoEnergyOverCpu{} and \llmTwoEnergyOverGpu{} compared to CPU and GPU implementations, respectively.
With 4-bit models, \MVDRAM{} maintains a \llmFourEnergyOverCpu{} energy efficiency advantage over CPU implementations.
These results highlight \MVDRAM{}'s superior energy efficiency when compared to CPU implementations using the same DDR4 DRAM technology.
 \section{Discussion}\label{sec:discussion}

\subsubsection*{\textbf{System Integration}}
\MVDRAM{} integrates with existing systems by requiring no modifications to DRAM while only needing PUD command support from the processor.
This integration involves enhancing the processor's memory controller to handle PUD-specific timing controls and providing a programming interface for applications to access these capabilities.
When the program requests to execute \RowCopy{} or \MAJ{X} operations through the interface, the memory controller will issue DRAM commands with specialized timing to perform the operations.
Both key techniques of \MVDRAM{}, on-the-fly vector encoding and horizontal matrix layout, can be implemented following this design.
Prior works~\cite{hajinazarSIMDRAMFrameworkBitserial2021,oliveiraMIMDRAMEndtoEndProcessingUsingDRAM2024} have proposed similar approaches with memory controller extensions and ISA-based programming interfaces.
These implementations demonstrate that the on-chip circuit area overhead, including their dedicated computation units, remains below 1\% of the total chip area. 

\subsubsection*{\textbf{Applicability to Other DRAM Technologies}}
While PUD operations have been experimentally verified only on DDR3 and DDR4 memory modules~\pudUnmodifiedCite{}, we expect that exposing PUD functionality in other DRAM standards would require only minimal modifications without subarray circuit changes.
Within the same memory vendor, the core subarray circuitry, which PUD directly leverages, remains largely consistent across different DRAM technologies, including DDR, LPDDR, and even HBM~\cite{namDRAMScopeUncoveringDRAM2024a}.
This architectural consistency stems from manufacturers' emphasis on maintaining proven subarray designs, which represent the most performance-critical and density-sensitive components of memory chips.
Based on this observation, we anticipate that the PUD operations demonstrated in DDR3 and DDR4 can be achieved in these alternative DRAM architectures without requiring technically challenging and costly subarray circuit modifications.
This suggests that \MVDRAM{}'s approach could potentially extend beyond current DDR4 implementations to future memory technologies and form factors, further expanding its applicability in diverse computing environments.

\subsubsection*{\textbf{Reliability}}
Due to the analog nature of PUD operations, \MVDRAM{} computations are susceptible to environmental factors including temperature variations, voltage fluctuations, and device aging.
However, prior research~\cite{yukselSimultaneousManyRowActivation2024} has evaluated the resilience of PUD's fundamental \RowCopy{} and \MAJ{X} operations against various environmental factors.
For temperature variations, experiments show only a 0.07\% average decrease in reliable column count for simultaneous many-row activation when temperature increases from 50 \textdegree$C$ to 90 \textdegree$C$.
Similarly, underscaling voltage from 2.5 $\si{V}$ to 2.1 $\si{V}$ results in at most a 0.41\% decrease in reliable column count.
These results indicate that \MVDRAM{}'s PUD operations are robust to environmental factors, ensuring reliable column count even under dynamic conditions.
 \section{Related Work}\label{sec:related}
To the best of our knowledge, \MVDRAM{} is the first system to accelerate GeMV operations for low-bit LLM inference using unmodified DRAM.
In this section, we discuss some of the closely related work.

\subsubsection*{\textbf{PUD with Unmodified DRAM}}
Several prior works~\pudUnmodifiedCite{} have demonstrated the PUD functionality on commercial DRAM modules without hardware modifications.
ComputeDRAM~\cite{gaoComputeDRAMInMemoryCompute2019} first experimentally verified the fundamental \RowCopy{} and \MAJ{X} operations on commercial DDR3 DRAM modules without hardware modifications.
Subsequent works have expanded the capabilities of PUD operations, including intermediate value storage~\cite{gaoFracDRAMFractionalValues2022}, true random number generation~\cite{olgunQUACTRNGHighThroughputTrue2021}, and further expanded boolean logic operations~\cite{yukselFunctionallyCompleteBooleanLogic2024,yukselSimultaneousManyRowActivation2024}.
However, these prior works have primarily demonstrated only a limited set of primitive operations.
In contrast, our work builds upon these fundamental operations to demonstrate application-level practicality by implementing complete GeMV functionality for low-bit LLM acceleration.

\subsubsection*{\textbf{PUD with Modified DRAM}}
While several works~\pudModifiedCite{} have proposed modifications to DRAM hardware to enhance PUD's computational capabilities, \MVDRAM{} addresses functional limitations through processor-DRAM co-design rather than circuit modifications.
Modifying DRAM subarray circuits presents significant technical and cost challenges due to nanoscale manufacturing processes, extremely dense designs, stringent performance requirements, and extensive validation needed for mass production~\cite{marazziHiFiDRAMEnablingHighfidelity2024,hoonProcessinginMemoryArchitectureCommodity2024}.
Some research efforts~\pudModifiedSmallCite{} have proposed minimal circuit modifications to enhance DRAM's computational capabilities, such as supporting \NOT{} operations or enabling data movement between subarrays.
Other proposals~\pudModifiedLargeCite{} introduce more substantial modifications, which would further complicate manufacturing processes and increase costs.

SIMDRAM~\cite{hajinazarSIMDRAMFrameworkBitserial2021} and MIMDRAM~\cite{oliveiraMIMDRAMEndtoEndProcessingUsingDRAM2024} have proposed PUD frameworks and evaluated matrix operations using these frameworks, involving various modifications to DRAM and specialized processing units on-chip.
Their evaluations primarily focus on the latency of matrix multiplication kernels in isolation.
In contrast, our work addresses end-to-end acceleration within an inference pipeline through processor-DRAM coordination while working within the constraints of existing commercial DRAM.

\subsubsection*{\textbf{Processing-near-DRAM}}
Processing-near-DRAM (PnD) approaches add computing logic near memory arrays, which are classified as a type of Processing-in-Memory (PIM)~\cite{mutluMemoryCentricComputingRecent2024}.
Commercial prototypes have been manufactured by major DRAM vendors~\cite{leeHardwareArchitectureSoftware2021,kwonSystemArchitectureSoftware2022}, while UPMEM has released commercially available PnD solutions~\cite{devauxTrueProcessingMemory2019}.
Several works~\pndCite{} have proposed methods to accelerate LLM inference using these PnD architectures.
While these approaches enable heterogeneous computing capabilities within memory systems, they generally exhibit lower area and power efficiency compared to PUD techniques, as they require dedicated processing units.

 \section{Conclusion}\label{sec:conclusion}
This paper presented \MVDRAM{}, the first system to realize matrix-vector multiplication for end-to-end low-bit LLM inference using unmodified DRAM.
Through on-the-fly vector encoding and horizontal matrix layout, \MVDRAM{} eliminates the overheads introduced by the PUD's fundamental limitation of column-to-column data movement.
Our evaluation on real DDR4 DRAM modules demonstrated up to \gemvLatencyOverCpu{} speedup and \gemvEnergyOverCpu{} energy efficiency for low-bit GeMV operations, with \llmTwoThroughputOverCpu{} throughput and \llmTwoEnergyOverCpu{} energy efficiency improvements for low-bit quantized models. 
\arxiv{
    \MVDRAM{} demonstrates that standard DRAM can serve not only as model storage but also as an inference accelerator.
    This dual-purpose approach has profound implications for resource-constrained devices, as it enables high-performance LLM inference without requiring specialized accelerators.
}

\bibliographystyle{IEEEtran}

\begin{thebibliography}{10}
\providecommand{\url}[1]{#1}
\csname url@samestyle\endcsname
\providecommand{\newblock}{\relax}
\providecommand{\bibinfo}[2]{#2}
\providecommand{\BIBentrySTDinterwordspacing}{\spaceskip=0pt\relax}
\providecommand{\BIBentryALTinterwordstretchfactor}{4}
\providecommand{\BIBentryALTinterwordspacing}{\spaceskip=\fontdimen2\font plus
\BIBentryALTinterwordstretchfactor\fontdimen3\font minus
  \fontdimen4\font\relax}
\providecommand{\BIBforeignlanguage}[2]{{\expandafter\ifx\csname l@#1\endcsname\relax
\typeout{** WARNING: IEEEtran.bst: No hyphenation pattern has been}\typeout{** loaded for the language `#1'. Using the pattern for}\typeout{** the default language instead.}\else
\language=\csname l@#1\endcsname
\fi
#2}}
\providecommand{\BIBdecl}{\relax}
\BIBdecl

\bibitem{gunterAppleIntelligenceFoundation2024}
T.~Gunter, Z.~Wang, C.~Wang, R.~Pang, A.~Narayanan, A.~Zhang, B.~Zhang,
  C.~Chen, C.-C. Chiu, D.~Qiu, D.~Gopinath, D.~A. Yap, D.~Yin, F.~Nan,
  F.~Weers, G.~Yin, H.~Huang, J.~Wang, J.~Lu, J.~Peebles, K.~Ye, M.~Lee, N.~Du,
  Q.~Chen, Q.~Keunebroek, S.~Wiseman, S.~Evans, T.~Lei, V.~Rathod, X.~Kong,
  X.~Du, Y.~Li, Y.~Wang, Y.~Gao, Z.~Ahmed, Z.~Xu, Z.~Lu, A.~Rashid, A.~M. Jose,
  A.~Doane, A.~Bencomo, A.~Vanderby, A.~Hansen, A.~Jain, A.~M. Anupama,
  A.~Kamal, B.~Wu, C.~Brum, C.~Maalouf, C.~Erdenebileg, C.~Dulhanty, D.~Moritz,
  D.~Kang, E.~Jimenez, E.~Ladd, F.~Shi, F.~Bai, F.~Chu, F.~Hohman, H.~Kotek,
  H.~G. Coleman, J.~Li, J.~Bigham, J.~Cao, J.~Lai, J.~Cheung, J.~Shan, J.~Zhou,
  J.~Li, J.~Qin, K.~Singh, K.~Vega, K.~Zou, L.~Heckman, L.~Gardiner, M.~Bowler,
  M.~Cordell, M.~Cao, N.~Hay, N.~Shahdadpuri, O.~Godwin, P.~Dighe,
  P.~Rachapudi, R.~Tantawi, R.~Frigg, S.~Davarnia, S.~Shah, S.~Guha,
  S.~Sirovica, S.~Ma, S.~Ma, S.~Wang, S.~Kim, S.~Jayaram, V.~Shankar, V.~Paidi,
  V.~Kumar, X.~Wang, X.~Zheng, W.~Cheng, Y.~Shrager, Y.~Ye, Y.~Tanaka, Y.~Guo,
  Y.~Meng, Z.~T. Luo, Z.~Ouyang, A.~Aygar, A.~Wan, A.~Walkingshaw,
  A.~Narayanan, A.~Lin, A.~Farooq, B.~Ramerth, C.~Reed, C.~Bartels, C.~Chaney,
  D.~Riazati, E.~L. Yang, E.~Feldman, G.~Hochstrasser, G.~Seguin, I.~Belousova,
  J.~Pelemans, K.~Yang, K.~A. Vahid, L.~Cao, M.~Najibi, M.~Zuliani, M.~Horton,
  M.~Cho, N.~Bhendawade, P.~Dong, P.~Maj, P.~Agrawal, Q.~Shan, Q.~Fu,
  R.~Poston, S.~Xu, S.~Liu, S.~Rao, T.~Heeramun, T.~Merth, U.~Rayala, V.~Cui,
  V.~R. Sridhar, W.~Zhang, W.~Zhang, W.~Wu, X.~Zhou, X.~Liu, Y.~Zhao, Y.~Xia,
  Z.~Ren, and Z.~Ren, ``Apple {{Intelligence Foundation Language Models}},''
  Jul. 2024.

\bibitem{pradeep2024phi}
\BIBentryALTinterwordspacing
V.~Pradeep, ``Phi silica, small but mighty on-device slm,'' December 2024,
  accessed: 2025-03-29. [Online]. Available:
  \url{https://blogs.windows.com/windowsexperience/2024/12/06/phi-silica-small-but-mighty-on-device-slm/}
\BIBentrySTDinterwordspacing

\bibitem{teamGeminiFamilyHighly2024}
G.~Team, R.~Anil, S.~Borgeaud, J.-B. Alayrac, J.~Yu, R.~Soricut, J.~Schalkwyk,
  A.~M. Dai, A.~Hauth, K.~Millican, D.~Silver, M.~Johnson, I.~Antonoglou,
  J.~Schrittwieser, A.~Glaese, J.~Chen, E.~Pitler, T.~Lillicrap, A.~Lazaridou,
  O.~Firat, J.~Molloy, M.~Isard, P.~R. Barham, T.~Hennigan, B.~Lee, F.~Viola,
  M.~Reynolds, Y.~Xu, R.~Doherty, E.~Collins, C.~Meyer, E.~Rutherford,
  E.~Moreira, K.~Ayoub, M.~Goel, J.~Krawczyk, C.~Du, E.~Chi, H.-T. Cheng,
  E.~Ni, P.~Shah, P.~Kane, B.~Chan, M.~Faruqui, A.~Severyn, H.~Lin, Y.~Li,
  Y.~Cheng, A.~Ittycheriah, M.~Mahdieh, M.~Chen, P.~Sun, D.~Tran, S.~Bagri,
  B.~Lakshminarayanan, J.~Liu, A.~Orban, F.~G{\"u}ra, H.~Zhou, X.~Song,
  A.~Boffy, H.~Ganapathy, S.~Zheng, H.~Choe, {\'A}.~Weisz, T.~Zhu, Y.~Lu,
  S.~Gopal, J.~Kahn, M.~Kula, J.~Pitman, R.~Shah, E.~Taropa, M.~A. Merey,
  M.~Baeuml, Z.~Chen, L.~E. Shafey, Y.~Zhang, O.~Sercinoglu, G.~Tucker,
  E.~Piqueras, M.~Krikun, I.~Barr, N.~Savinov, I.~Danihelka, B.~Roelofs,
  A.~White, A.~Andreassen, T.~von Glehn, L.~Yagati, M.~Kazemi, L.~Gonzalez,
  M.~Khalman, J.~Sygnowski, A.~Frechette, C.~Smith, L.~Culp, L.~Proleev,
  Y.~Luan, X.~Chen, J.~Lottes, N.~Schucher, F.~Lebron, A.~Rrustemi, N.~Clay,
  P.~Crone, T.~Kocisky, J.~Zhao, B.~Perz, D.~Yu, H.~Howard, A.~Bloniarz, J.~W.
  Rae, H.~Lu, L.~Sifre, M.~Maggioni, F.~Alcober, D.~Garrette, M.~Barnes,
  S.~Thakoor, J.~Austin, G.~{Barth-Maron}, W.~Wong, R.~Joshi, R.~Chaabouni,
  D.~Fatiha, A.~Ahuja, G.~S. Tomar, E.~Senter, M.~Chadwick, I.~Kornakov,
  N.~Attaluri, I.~Iturrate, R.~Liu, Y.~Li, S.~Cogan, J.~Chen, C.~Jia, C.~Gu,
  Q.~Zhang, J.~Grimstad, A.~J. Hartman, X.~Garcia, T.~S. Pillai, J.~Devlin,
  M.~Laskin, D.~d.~L. Casas, D.~Valter, C.~Tao, L.~Blanco, A.~P. Badia,
  D.~Reitter, M.~Chen, J.~Brennan, C.~Rivera, S.~Brin, S.~Iqbal, G.~Surita,
  J.~Labanowski, A.~Rao, S.~Winkler, E.~Parisotto, Y.~Gu, K.~Olszewska,
  R.~Addanki, A.~Miech, A.~Louis, D.~Teplyashin, G.~Brown, E.~Catt,
  J.~Balaguer, J.~Xiang, P.~Wang, Z.~Ashwood, A.~Briukhov, A.~Webson,
  S.~Ganapathy, S.~Sanghavi, A.~Kannan, M.-W. Chang, A.~Stjerngren,
  J.~Djolonga, Y.~Sun, A.~Bapna, M.~Aitchison, P.~Pejman, H.~Michalewski,
  T.~Yu, C.~Wang, J.~Love, J.~Ahn, D.~Bloxwich, K.~Han, P.~Humphreys,
  T.~Sellam, J.~Bradbury, V.~Godbole, S.~Samangooei, B.~Damoc, A.~Kaskasoli,
  S.~M.~R. Arnold, V.~Vasudevan, S.~Agrawal, J.~Riesa, D.~Lepikhin, R.~Tanburn,
  S.~Srinivasan, H.~Lim, S.~Hodkinson, P.~Shyam, J.~Ferret, S.~Hand, A.~Garg,
  T.~L. Paine, J.~Li, Y.~Li, M.~Giang, A.~Neitz, Z.~Abbas, S.~York, M.~Reid,
  E.~Cole, A.~Chowdhery, D.~Das, D.~Rogozi{\'n}ska, V.~Nikolaev, P.~Sprechmann,
  Z.~Nado, L.~Zilka, F.~Prost, L.~He, M.~Monteiro, G.~Mishra, C.~Welty,
  J.~Newlan, D.~Jia, M.~Allamanis, C.~H. Hu, R.~de~Liedekerke, J.~Gilmer,
  C.~Saroufim, S.~Rijhwani, S.~Hou, D.~Shrivastava, A.~Baddepudi, A.~Goldin,
  A.~Ozturel, A.~Cassirer, Y.~Xu, D.~Sohn, D.~Sachan, R.~K. Amplayo,
  C.~Swanson, D.~Petrova, S.~Narayan, A.~Guez, S.~Brahma, J.~Landon, M.~Patel,
  R.~Zhao, K.~Villela, L.~Wang, W.~Jia, M.~Rahtz, M.~Gim{\'e}nez, L.~Yeung,
  J.~Keeling, P.~Georgiev, D.~Mincu, B.~Wu, S.~Haykal, R.~Saputro,
  K.~Vodrahalli, J.~Qin, Z.~Cankara, A.~Sharma, N.~Fernando, W.~Hawkins,
  B.~Neyshabur, S.~Kim, A.~Hutter, P.~Agrawal, A.~{Castro-Ros}, G.~van~den
  Driessche, T.~Wang, F.~Yang, S.-y. Chang, P.~Komarek, R.~McIlroy, M.~Lu{\v
  c}i{\'c}, G.~Zhang, W.~Farhan, M.~Sharman, P.~Natsev, P.~Michel, Y.~Bansal,
  S.~Qiao, K.~Cao, S.~Shakeri, C.~Butterfield, J.~Chung, P.~K. Rubenstein,
  S.~Agrawal, A.~Mensch, K.~Soparkar, K.~Lenc, T.~Chung, A.~Pope, L.~Maggiore,
  J.~Kay, P.~Jhakra, S.~Wang, J.~Maynez, M.~Phuong, T.~Tobin, A.~Tacchetti,
  M.~Trebacz, K.~Robinson, Y.~Katariya, S.~Riedel, P.~Bailey, K.~Xiao,
  N.~Ghelani, L.~Aroyo, A.~Slone, N.~Houlsby, X.~Xiong, Z.~Yang,
  E.~Gribovskaya, J.~Adler, M.~Wirth, L.~Lee, M.~Li, T.~Kagohara, J.~Pavagadhi,
  S.~Bridgers, A.~Bortsova, S.~Ghemawat, Z.~Ahmed, T.~Liu, R.~Powell,
  V.~Bolina, M.~Iinuma, P.~Zablotskaia, J.~Besley, D.-W. Chung, T.~Dozat,
  R.~Comanescu, X.~Si, J.~Greer, G.~Su, M.~Polacek, R.~L. Kaufman, S.~Tokumine,
  H.~Hu, E.~Buchatskaya, Y.~Miao, M.~Elhawaty, A.~Siddhant, N.~Tomasev,
  J.~Xing, C.~Greer, H.~Miller, S.~Ashraf, A.~Roy, Z.~Zhang, A.~Ma, A.~Filos,
  M.~Besta, R.~Blevins, T.~Klimenko, C.-K. Yeh, S.~Changpinyo, J.~Mu, O.~Chang,
  M.~Pajarskas, C.~Muir, V.~Cohen, C.~L. Lan, K.~Haridasan, A.~Marathe,
  S.~Hansen, S.~Douglas, R.~Samuel, M.~Wang, S.~Austin, C.~Lan, J.~Jiang,
  J.~Chiu, J.~A. Lorenzo, L.~L. Sj{\"o}sund, S.~Cevey, Z.~Gleicher,
  T.~Avrahami, A.~Boral, H.~Srinivasan, V.~Selo, R.~May, K.~Aisopos,
  L.~Hussenot, L.~B. Soares, K.~Baumli, M.~B. Chang, A.~Recasens, B.~Caine,
  A.~Pritzel, F.~Pavetic, F.~Pardo, A.~Gergely, J.~Frye, V.~Ramasesh,
  D.~Horgan, K.~Badola, N.~Kassner, S.~Roy, E.~Dyer, V.~C. Campos, A.~Tomala,
  Y.~Tang, D.~E. Badawy, E.~White, B.~Mustafa, O.~Lang, A.~Jindal, S.~Vikram,
  Z.~Gong, S.~Caelles, R.~Hemsley, G.~Thornton, F.~Feng, W.~Stokowiec,
  C.~Zheng, P.~Thacker, {\c C}.~{\"U}nl{\"u}, Z.~Zhang, M.~Saleh, J.~Svensson,
  M.~Bileschi, P.~Patil, A.~Anand, R.~Ring, K.~Tsihlas, A.~Vezer, M.~Selvi,
  T.~Shevlane, M.~Rodriguez, T.~Kwiatkowski, S.~Daruki, K.~Rong, A.~Dafoe,
  N.~FitzGerald, K.~{Gu-Lemberg}, M.~Khan, L.~A. Hendricks, M.~Pellat,
  V.~Feinberg, J.~{Cobon-Kerr}, T.~Sainath, M.~Rauh, S.~H. Hashemi, R.~Ives,
  Y.~Hasson, E.~Noland, Y.~Cao, N.~Byrd, L.~Hou, Q.~Wang, T.~Sottiaux,
  M.~Paganini, J.-B. Lespiau, A.~Moufarek, S.~Hassan, K.~Shivakumar, J.~van
  Amersfoort, A.~Mandhane, P.~Joshi, A.~Goyal, M.~Tung, A.~Brock, H.~Sheahan,
  V.~Misra, C.~Li, N.~Raki{\'c}evi{\'c}, M.~Dehghani, F.~Liu, S.~Mittal, J.~Oh,
  S.~Noury, E.~Sezener, F.~Huot, M.~Lamm, N.~D. Cao, C.~Chen, S.~Mudgal,
  R.~Stella, K.~Brooks, G.~Vasudevan, C.~Liu, M.~Chain, N.~Melinkeri, A.~Cohen,
  V.~Wang, K.~Seymore, S.~Zubkov, R.~Goel, S.~Yue, S.~Krishnakumaran,
  B.~Albert, N.~Hurley, M.~Sano, A.~Mohananey, J.~Joughin, E.~Filonov, T.~K{\k
  e}pa, Y.~Eldawy, J.~Lim, R.~Rishi, S.~Badiezadegan, T.~Bos, J.~Chang,
  S.~Jain, S.~G.~S. Padmanabhan, S.~Puttagunta, K.~Krishna, L.~Baker, N.~Kalb,
  V.~Bedapudi, A.~Kurzrok, S.~Lei, A.~Yu, O.~Litvin, X.~Zhou, Z.~Wu, S.~Sobell,
  A.~Siciliano, A.~Papir, R.~Neale, J.~Bragagnolo, T.~Toor, T.~Chen, V.~Anklin,
  F.~Wang, R.~Feng, M.~Gholami, K.~Ling, L.~Liu, J.~Walter, H.~Moghaddam,
  A.~Kishore, J.~Adamek, T.~Mercado, J.~Mallinson, S.~Wandekar, S.~Cagle,
  E.~Ofek, G.~Garrido, C.~Lombriser, M.~Mukha, B.~Sun, H.~R. Mohammad,
  J.~Matak, Y.~Qian, V.~Peswani, P.~Janus, Q.~Yuan, L.~Schelin, O.~David,
  A.~Garg, Y.~He, O.~Duzhyi, A.~{\"A}lgmyr, T.~Lottaz, Q.~Li, V.~Yadav, L.~Xu,
  A.~Chinien, R.~Shivanna, A.~Chuklin, J.~Li, C.~Spadine, T.~Wolfe, K.~Mohamed,
  S.~Das, Z.~Dai, K.~He, D.~von Dincklage, S.~Upadhyay, A.~Maurya, L.~Chi,
  S.~Krause, K.~Salama, P.~G. Rabinovitch, P.~K.~R. M, A.~Selvan, M.~Dektiarev,
  G.~Ghiasi, E.~Guven, H.~Gupta, B.~Liu, D.~Sharma, I.~H. Shtacher, S.~Paul,
  O.~Akerlund, F.-X. Aubet, T.~Huang, C.~Zhu, E.~Zhu, E.~Teixeira, M.~Fritze,
  F.~Bertolini, L.-E. Marinescu, M.~B{\"o}lle, D.~Paulus, K.~Gupta, T.~Latkar,
  M.~Chang, J.~Sanders, R.~Wilson, X.~Wu, Y.-X. Tan, L.~N. Thiet, T.~Doshi,
  S.~Lall, S.~Mishra, W.~Chen, T.~Luong, S.~Benjamin, J.~Lee, E.~Andrejczuk,
  D.~Rabiej, V.~Ranjan, K.~Styrc, P.~Yin, J.~Simon, M.~R. Harriott, M.~Bansal,
  A.~Robsky, G.~Bacon, D.~Greene, D.~Mirylenka, C.~Zhou, O.~Sarvana, A.~Goyal,
  S.~Andermatt, P.~Siegler, B.~Horn, A.~Israel, F.~Pongetti, C.-W.~L. Chen,
  M.~Selvatici, P.~Silva, K.~Wang, J.~Tolins, K.~Guu, R.~Yogev, X.~Cai,
  A.~Agostini, M.~Shah, H.~Nguyen, N.~{\'O}. Donnaile, S.~Pereira, L.~Friso,
  A.~Stambler, A.~Kurzrok, C.~Kuang, Y.~Romanikhin, M.~Geller, Z.~J. Yan,
  K.~Jang, C.-C. Lee, W.~Fica, E.~Malmi, Q.~Tan, D.~Banica, D.~Balle, R.~Pham,
  Y.~Huang, D.~Avram, H.~Shi, J.~Singh, C.~Hidey, N.~Ahuja, P.~Saxena,
  D.~Dooley, S.~P. Potharaju, E.~O'Neill, A.~Gokulchandran, R.~Foley, K.~Zhao,
  M.~Dusenberry, Y.~Liu, P.~Mehta, R.~Kotikalapudi, C.~{Safranek-Shrader},
  A.~Goodman, J.~Kessinger, E.~Globen, P.~Kolhar, C.~Gorgolewski, A.~Ibrahim,
  Y.~Song, A.~Eichenbaum, T.~Brovelli, S.~Potluri, P.~Lahoti, C.~Baetu,
  A.~Ghorbani, C.~Chen, A.~Crawford, S.~Pal, M.~Sridhar, P.~Gurita, A.~Mujika,
  I.~Petrovski, P.-L. Cedoz, C.~Li, S.~Chen, N.~D. Santo, S.~Goyal, J.~Punjabi,
  K.~Kappaganthu, C.~Kwak, P.~LV, S.~Velury, H.~Choudhury, J.~Hall, P.~Shah,
  R.~Figueira, M.~Thomas, M.~Lu, T.~Zhou, C.~Kumar, T.~Jurdi, S.~Chikkerur,
  Y.~Ma, A.~Yu, S.~Kwak, V.~{\"A}hdel, S.~Rajayogam, T.~Choma, F.~Liu,
  A.~Barua, C.~Ji, J.~H. Park, V.~Hellendoorn, A.~Bailey, T.~Bilal, H.~Zhou,
  M.~Khatir, C.~Sutton, W.~Rzadkowski, F.~Macintosh, K.~Shagin, P.~Medina,
  C.~Liang, J.~Zhou, P.~Shah, Y.~Bi, A.~Dankovics, S.~Banga, S.~Lehmann,
  M.~Bredesen, Z.~Lin, J.~E. Hoffmann, J.~Lai, R.~Chung, K.~Yang, N.~Balani,
  A.~Bra{\v z}inskas, A.~Sozanschi, M.~Hayes, H.~F. Alcalde, P.~Makarov,
  W.~Chen, A.~Stella, L.~Snijders, M.~Mandl, A.~K{\"a}rrman, P.~Nowak, X.~Wu,
  A.~Dyck, K.~Vaidyanathan, R.~R, J.~Mallet, M.~Rudominer, E.~Johnston,
  S.~Mittal, A.~Udathu, J.~Christensen, V.~Verma, Z.~Irving, A.~Santucci,
  G.~Elsayed, E.~Davoodi, M.~Georgiev, I.~Tenney, N.~Hua, G.~Cideron,
  E.~Leurent, M.~Alnahlawi, I.~Georgescu, N.~Wei, I.~Zheng, D.~Scandinaro,
  H.~Jiang, J.~Snoek, M.~Sundararajan, X.~Wang, Z.~Ontiveros, I.~Karo, J.~Cole,
  V.~Rajashekhar, L.~Tumeh, E.~{Ben-David}, R.~Jain, J.~Uesato, R.~Datta,
  O.~Bunyan, S.~Wu, J.~Zhang, P.~Stanczyk, Y.~Zhang, D.~Steiner, S.~Naskar,
  M.~Azzam, M.~Johnson, A.~Paszke, C.-C. Chiu, J.~S. Elias, A.~Mohiuddin,
  F.~Muhammad, J.~Miao, A.~Lee, N.~Vieillard, J.~Park, J.~Zhang, J.~Stanway,
  D.~Garmon, A.~Karmarkar, Z.~Dong, J.~Lee, A.~Kumar, L.~Zhou, J.~Evens,
  W.~Isaac, G.~Irving, E.~Loper, M.~Fink, I.~Arkatkar, N.~Chen, I.~Shafran,
  I.~Petrychenko, Z.~Chen, J.~Jia, A.~Levskaya, Z.~Zhu, P.~Grabowski, Y.~Mao,
  A.~Magni, K.~Yao, J.~Snaider, N.~Casagrande, E.~Palmer, P.~Suganthan,
  A.~Casta{\~n}o, I.~Giannoumis, W.~Kim, M.~Rybi{\'n}ski, A.~Sreevatsa,
  J.~Prendki, D.~Soergel, A.~Goedeckemeyer, W.~Gierke, M.~Jafari, M.~Gaba,
  J.~Wiesner, D.~G. Wright, Y.~Wei, H.~Vashisht, Y.~Kulizhskaya, J.~Hoover,
  M.~Le, L.~Li, C.~Iwuanyanwu, L.~Liu, K.~Ramirez, A.~Khorlin, A.~Cui, T.~LIN,
  M.~Wu, R.~Aguilar, K.~Pallo, A.~Chakladar, G.~Perng, E.~A. Abellan, M.~Zhang,
  I.~Dasgupta, N.~Kushman, I.~Penchev, A.~Repina, X.~Wu, T.~van~der Weide,
  P.~Ponnapalli, C.~Kaplan, J.~Simsa, S.~Li, O.~Dousse, F.~Yang, J.~Piper,
  N.~Ie, R.~Pasumarthi, N.~Lintz, A.~Vijayakumar, D.~Andor, P.~Valenzuela,
  M.~Lui, C.~Paduraru, D.~Peng, K.~Lee, S.~Zhang, S.~Greene, D.~D. Nguyen,
  P.~Kurylowicz, C.~Hardin, L.~Dixon, L.~Janzer, K.~Choo, Z.~Feng, B.~Zhang,
  A.~Singhal, D.~Du, D.~McKinnon, N.~Antropova, T.~Bolukbasi, O.~Keller,
  D.~Reid, D.~Finchelstein, M.~A. Raad, R.~Crocker, P.~Hawkins, R.~Dadashi,
  C.~Gaffney, K.~Franko, A.~Bulanova, R.~Leblond, S.~Chung, H.~Askham, L.~C.
  Cobo, K.~Xu, F.~Fischer, J.~Xu, C.~Sorokin, C.~Alberti, C.-C. Lin, C.~Evans,
  A.~Dimitriev, H.~Forbes, D.~Banarse, Z.~Tung, M.~Omernick, C.~Bishop,
  R.~Sterneck, R.~Jain, J.~Xia, E.~Amid, F.~Piccinno, X.~Wang, P.~Banzal, D.~J.
  Mankowitz, A.~Polozov, V.~Krakovna, S.~Brown, M.~Bateni, D.~Duan, V.~Firoiu,
  M.~Thotakuri, T.~Natan, M.~Geist, S.~tan Girgin, H.~Li, J.~Ye, O.~Roval,
  R.~Tojo, M.~Kwong, J.~{Lee-Thorp}, C.~Yew, D.~Sinopalnikov, S.~Ramos,
  J.~Mellor, A.~Sharma, K.~Wu, D.~Miller, N.~Sonnerat, D.~Vnukov, R.~Greig,
  J.~Beattie, E.~Caveness, L.~Bai, J.~Eisenschlos, A.~Korchemniy, T.~Tsai,
  M.~Jasarevic, W.~Kong, P.~Dao, Z.~Zheng, F.~Liu, F.~Yang, R.~Zhu, T.~H. Teh,
  J.~Sanmiya, E.~Gladchenko, N.~Trdin, D.~Toyama, E.~Rosen, S.~Tavakkol,
  L.~Xue, C.~Elkind, O.~Woodman, J.~Carpenter, G.~Papamakarios, R.~Kemp,
  S.~Kafle, T.~Grunina, R.~Sinha, A.~Talbert, D.~Wu, D.~{Owusu-Afriyie}, C.~Du,
  C.~Thornton, J.~{Pont-Tuset}, P.~Narayana, J.~Li, S.~Fatehi, J.~Wieting,
  O.~Ajmeri, B.~Uria, Y.~Ko, L.~Knight, A.~H{\'e}liou, N.~Niu, S.~Gu, C.~Pang,
  Y.~Li, N.~Levine, A.~Stolovich, R.~{Santamaria-Fernandez}, S.~Goenka,
  W.~Yustalim, R.~Strudel, A.~Elqursh, C.~Deck, H.~Lee, Z.~Li, K.~Levin,
  R.~Hoffmann, D.~{Holtmann-Rice}, O.~Bachem, S.~Arora, C.~Koh, S.~H. Yeganeh,
  S.~P{\~o}der, M.~Tariq, Y.~Sun, L.~Ionita, M.~Seyedhosseini, P.~Tafti,
  Z.~Liu, A.~Gulati, J.~Liu, X.~Ye, B.~Chrzaszcz, L.~Wang, N.~Sethi, T.~Li,
  B.~Brown, S.~Singh, W.~Fan, A.~Parisi, J.~Stanton, V.~Koverkathu, C.~A.
  {Choquette-Choo}, Y.~Li, T.~J. Lu, A.~Ittycheriah, P.~Shroff, M.~Varadarajan,
  S.~Bahargam, R.~Willoughby, D.~Gaddy, G.~Desjardins, M.~Cornero, B.~Robenek,
  B.~Mittal, B.~Albrecht, A.~Shenoy, F.~Moiseev, H.~Jacobsson, A.~Ghaffarkhah,
  M.~Rivi{\`e}re, A.~Walton, C.~Crepy, A.~Parrish, Z.~Zhou, C.~Farabet,
  C.~Radebaugh, P.~Srinivasan, C.~van~der Salm, A.~Fidjeland, S.~Scellato,
  E.~{Latorre-Chimoto}, H.~{Klimczak-Pluci{\'n}ska}, D.~Bridson, D.~de~Cesare,
  T.~Hudson, P.~Mendolicchio, L.~Walker, A.~Morris, M.~Mauger, A.~Guseynov,
  A.~Reid, S.~Odoom, L.~Loher, V.~Cotruta, M.~Yenugula, D.~Grewe,
  A.~Petrushkina, T.~Duerig, A.~Sanchez, S.~Yadlowsky, A.~Shen, A.~Globerson,
  L.~Webb, S.~Dua, D.~Li, S.~Bhupatiraju, D.~Hurt, H.~Qureshi, A.~Agarwal,
  T.~Shani, M.~Eyal, A.~Khare, S.~R. Belle, L.~Wang, C.~Tekur, M.~S. Kale,
  J.~Wei, R.~Sang, B.~Saeta, T.~Liechty, Y.~Sun, Y.~Zhao, S.~Lee, P.~Nayak,
  D.~Fritz, M.~R. Vuyyuru, J.~Aslanides, N.~Vyas, M.~Wicke, X.~Ma, E.~Eltyshev,
  N.~Martin, H.~Cate, J.~Manyika, K.~Amiri, Y.~Kim, X.~Xiong, K.~Kang,
  F.~Luisier, N.~Tripuraneni, D.~Madras, M.~Guo, A.~Waters, O.~Wang,
  J.~Ainslie, J.~Baldridge, H.~Zhang, G.~Pruthi, J.~Bauer, F.~Yang, R.~Mansour,
  J.~Gelman, Y.~Xu, G.~Polovets, J.~Liu, H.~Cai, W.~Chen, X.~Sheng, E.~Xue,
  S.~Ozair, C.~Angermueller, X.~Li, A.~Sinha, W.~Wang, J.~Wiesinger,
  E.~Koukoumidis, Y.~Tian, A.~Iyer, M.~Gurumurthy, M.~Goldenson, P.~Shah, M.~K.
  Blake, H.~Yu, A.~Urbanowicz, J.~Palomaki, C.~Fernando, K.~Durden, H.~Mehta,
  N.~Momchev, E.~Rahimtoroghi, M.~Georgaki, A.~Raul, S.~Ruder, M.~Redshaw,
  J.~Lee, D.~Zhou, K.~Jalan, D.~Li, B.~Hechtman, P.~Schuh, M.~Nasr, K.~Milan,
  V.~Mikulik, J.~Franco, T.~Green, N.~Nguyen, J.~Kelley, A.~Mahendru, A.~Hu,
  J.~Howland, B.~Vargas, J.~Hui, K.~Bansal, V.~Rao, R.~Ghiya, E.~Wang, K.~Ye,
  J.~M. Sarr, M.~M. Preston, M.~Elish, S.~Li, A.~Kaku, J.~Gupta, I.~Pasupat,
  D.-C. Juan, M.~Someswar, T.~M, X.~Chen, A.~Amini, A.~Fabrikant, E.~Chu,
  X.~Dong, A.~Muthal, S.~Buthpitiya, S.~Jauhari, N.~Hua, U.~Khandelwal,
  A.~Hitron, J.~Ren, L.~Rinaldi, S.~Drath, A.~Dabush, N.-J. Jiang, H.~Godhia,
  U.~Sachs, A.~Chen, Y.~Fan, H.~Taitelbaum, H.~Noga, Z.~Dai, J.~Wang, C.~Liang,
  J.~Hamer, C.-S. Ferng, C.~Elkind, A.~Atias, P.~Lee, V.~List{\'i}k, M.~Carlen,
  J.~van~de Kerkhof, M.~Pikus, K.~Zaher, P.~M{\"u}ller, S.~Zykova, R.~Stefanec,
  V.~Gatsko, C.~Hirnschall, A.~Sethi, X.~F. Xu, C.~Ahuja, B.~Tsai,
  A.~Stefanoiu, B.~Feng, K.~Dhandhania, M.~Katyal, A.~Gupta, A.~Parulekar,
  D.~Pitta, J.~Zhao, V.~Bhatia, Y.~Bhavnani, O.~Alhadlaq, X.~Li, P.~Danenberg,
  D.~Tu, A.~Pine, V.~Filippova, A.~Ghosh, B.~Limonchik, B.~Urala, C.~K. Lanka,
  D.~Clive, Y.~Sun, E.~Li, H.~Wu, K.~Hongtongsak, I.~Li, K.~Thakkar, K.~Omarov,
  K.~Majmundar, M.~Alverson, M.~Kucharski, M.~Patel, M.~Jain, M.~Zabelin,
  P.~Pelagatti, R.~Kohli, S.~Kumar, J.~Kim, S.~Sankar, V.~Shah,
  L.~Ramachandruni, X.~Zeng, B.~Bariach, L.~Weidinger, T.~Vu, A.~Andreev,
  A.~He, K.~Hui, S.~Kashem, A.~Subramanya, S.~Hsiao, D.~Hassabis,
  K.~Kavukcuoglu, A.~Sadovsky, Q.~Le, T.~Strohman, Y.~Wu, S.~Petrov, J.~Dean,
  and O.~Vinyals, ``Gemini: {{A Family}} of {{Highly Capable Multimodal
  Models}},'' Jun. 2024.

\bibitem{kwonLoLPIMLongContextLLM2025}
H.~Kwon, K.~Koo, J.~Kim, W.~Lee, M.~Lee, H.~Lee, Y.~Jung, J.~Park, Y.~Song,
  B.~Yang, H.~Choi, G.~Kim, J.~Won, W.~Shin, C.~Kim, G.~Shin, Y.~Kwon, I.~Kim,
  E.~Lim, J.~Kim, and J.~Choi, ``{{LoL-PIM}}: {{Long-Context LLM Decoding}}
  with {{Scalable DRAM-PIM System}},'' Jan. 2025.

\bibitem{parkAttAccUnleashingPower2024}
J.~Park, J.~Choi, K.~Kyung, M.~J. Kim, Y.~Kwon, N.~S. Kim, and J.~H. Ahn,
  ``{{AttAcc}}! {{Unleashing}} the {{Power}} of {{PIM}} for {{Batched
  Transformer-based Generative Model Inference}},'' in \emph{Proceedings of the
  29th {{ACM International Conference}} on {{Architectural Support}} for
  {{Programming Languages}} and {{Operating Systems}}, {{Volume}} 2}, ser.
  {{ASPLOS}} '24, vol.~2.\hskip 1em plus 0.5em minus 0.4em\relax New York, NY,
  USA: Association for Computing Machinery, Apr. 2024, pp. 103--119.

\bibitem{hePAPIExploitingDynamic2025}
Y.~He, H.~Mao, C.~Giannoula, M.~Sadrosadati, J.~{G{\'o}mez-Luna}, H.~Li, X.~Li,
  Y.~Wang, and O.~Mutlu, ``{{PAPI}}: {{Exploiting Dynamic Parallelism}} in
  {{Large Language Model Decoding}} with a {{Processing-In-Memory-Enabled
  Computing System}},'' Feb. 2025.

\bibitem{seoIANUSIntegratedAccelerator2024}
M.~Seo, X.~T. Nguyen, S.~J. Hwang, Y.~Kwon, G.~Kim, C.~Park, I.~Kim, J.~Park,
  J.~Kim, W.~Shin, J.~Won, H.~Choi, K.~Kim, D.~Kwon, C.~Jeong, S.~Lee, Y.~Choi,
  W.~Byun, S.~Baek, H.-J. Lee, and J.~Kim, ``{{IANUS}}: {{Integrated
  Accelerator}} based on {{NPU-PIM Unified Memory System}},'' in
  \emph{Proceedings of the 29th {{ACM International Conference}} on
  {{Architectural Support}} for {{Programming Languages}} and {{Operating
  Systems}}, {{Volume}} 3}, ser. {{ASPLOS}} '24, vol.~3.\hskip 1em plus 0.5em
  minus 0.4em\relax New York, NY, USA: Association for Computing Machinery,
  Apr. 2024, pp. 545--560.

\bibitem{heoNeuPIMsNPUPIMHeterogeneous2024}
G.~Heo, S.~Lee, J.~Cho, H.~Choi, S.~Lee, H.~Ham, G.~Kim, D.~Mahajan, and
  J.~Park, ``{{NeuPIMs}}: {{NPU-PIM Heterogeneous Acceleration}} for {{Batched
  LLM Inferencing}},'' in \emph{Proceedings of the 29th {{ACM International
  Conference}} on {{Architectural Support}} for {{Programming Languages}} and
  {{Operating Systems}}, {{Volume}} 3}.\hskip 1em plus 0.5em minus 0.4em\relax
  La Jolla CA USA: ACM, Apr. 2024, pp. 722--737.

\bibitem{chengCompressedChainThought2024}
J.~Cheng and B.~V. Durme, ``Compressed {{Chain}} of {{Thought}}: {{Efficient
  Reasoning Through Dense Representations}},'' Dec. 2024.

\bibitem{ashkboosQuaRotOutlierFree4Bit2024}
S.~Ashkboos, A.~Mohtashami, M.~L. Croci, B.~Li, P.~Cameron, M.~Jaggi,
  D.~Alistarh, T.~Hoefler, and J.~Hensman, ``{{QuaRot}}: {{Outlier-Free}}
  4-{{Bit Inference}} in {{Rotated LLMs}},'' Oct. 2024.

\bibitem{zhao2024atom}
Y.~Zhao, C.-Y. Lin, K.~Zhu, Z.~Ye, L.~Chen, S.~Zheng, L.~Ceze,
  A.~Krishnamurthy, T.~Chen, and B.~Kasikci, ``Atom: Low-bit quantization for
  efficient and accurate llm serving,'' \emph{Proceedings of Machine Learning
  and Systems}, vol.~6, pp. 196--209, 2024.

\bibitem{elangovanBCQBlockClustered2025}
R.~Elangovan, C.~Sakr, A.~Raghunathan, and B.~Khailany, ``{{BCQ}}: {{Block
  Clustered Quantization}} for 4-bit ({{W4A4}}) {{LLM Inference}},'' Feb. 2025.

\bibitem{wangBitNetScaling1bit2023}
H.~Wang, S.~Ma, L.~Dong, S.~Huang, H.~Wang, L.~Ma, F.~Yang, R.~Wang, Y.~Wu, and
  F.~Wei, ``{{BitNet}}: {{Scaling}} 1-bit {{Transformers}} for {{Large Language
  Models}},'' Oct. 2023.

\bibitem{duBitDistillerUnleashingPotential2024}
D.~Du, Y.~Zhang, S.~Cao, J.~Guo, T.~Cao, X.~Chu, and N.~Xu, ``{{BitDistiller}}:
  {{Unleashing}} the {{Potential}} of {{Sub-4-Bit LLMs}} via
  {{Self-Distillation}},'' Feb. 2024.

\bibitem{chee2023quip}
J.~Chee, Y.~Cai, V.~Kuleshov, and C.~M. De~Sa, ``Quip: 2-bit quantization of
  large language models with guarantees,'' \emph{Advances in Neural Information
  Processing Systems}, vol.~36, pp. 4396--4429, 2023.

\bibitem{tsengQuIPEvenBetter2024}
A.~Tseng, J.~Chee, Q.~Sun, V.~Kuleshov, and C.~D. Sa, ``{{QuIP}}\#: {{Even
  Better LLM Quantization}} with {{Hadamard Incoherence}} and {{Lattice
  Codebooks}},'' Jun. 2024.

\bibitem{vptq}
Y.~Liu, J.~Wen, Y.~Wang, S.~Ye, L.~L. Zhang, T.~Cao, C.~Li, and M.~Yang,
  ``{VPTQ:} extreme low-bit vector post-training quantization for large
  language models,'' in \emph{Proceedings of the 2024 Conference on Empirical
  Methods in Natural Language Processing, {EMNLP} 2024, Miami, FL, USA,
  November 12-16, 2024}, Y.~Al{-}Onaizan, M.~Bansal, and Y.~Chen, Eds.\hskip
  1em plus 0.5em minus 0.4em\relax Association for Computational Linguistics,
  2024, pp. 8181--8196.

\bibitem{linQServeW4A8KV4Quantization2024}
Y.~Lin, H.~Tang, S.~Yang, Z.~Zhang, G.~Xiao, C.~Gan, and S.~Han, ``{{QServe}}:
  {{W4A8KV4 Quantization}} and {{System Co-design}} for {{Efficient LLM
  Serving}},'' May 2024.

\bibitem{xuOneBitExtremelyLowbit2024}
Y.~Xu, X.~Han, Z.~Yang, S.~Wang, Q.~Zhu, Z.~Liu, W.~Liu, and W.~Che,
  ``{{OneBit}}: {{Towards Extremely Low-bit Large Language Models}},'' Nov.
  2024.

\bibitem{wuUnderstandingINT4Quantization2023}
X.~Wu, C.~Li, R.~Y. Aminabadi, Z.~Yao, and Y.~He, ``Understanding {{INT4
  Quantization}} for {{Transformer Models}}: {{Latency Speedup}},
  {{Composability}}, and {{Failure Cases}},'' May 2023.

\bibitem{mutluMemoryCentricComputingRecent2024}
O.~Mutlu, A.~Olgun, G.~F. Oliveira, and I.~E. Yuksel, ``Memory-{{Centric
  Computing}}: {{Recent Advances}} in {{Processing-in-DRAM}},'' Dec. 2024.

\bibitem{seshadriRowCloneFastEnergyefficient2013}
V.~Seshadri, Y.~Kim, C.~Fallin, D.~Lee, R.~Ausavarungnirun, G.~Pekhimenko,
  Y.~Luo, O.~Mutlu, P.~B. Gibbons, M.~A. Kozuch, and T.~C. Mowry,
  ``{{RowClone}}: Fast and energy-efficient in-{{DRAM}} bulk data copy and
  initialization,'' in \emph{Proceedings of the 46th {{Annual IEEE}}/{{ACM
  International Symposium}} on {{Microarchitecture}}}.\hskip 1em plus 0.5em
  minus 0.4em\relax Davis California: ACM, Dec. 2013, pp. 185--197.

\bibitem{seshadriAmbitInmemoryAccelerator2017}
V.~Seshadri, D.~Lee, T.~Mullins, H.~Hassan, A.~Boroumand, J.~Kim, M.~A. Kozuch,
  O.~Mutlu, P.~B. Gibbons, and T.~C. Mowry, ``Ambit: In-memory accelerator for
  bulk bitwise operations using commodity {{DRAM}} technology,'' in
  \emph{Proceedings of the 50th {{Annual IEEE}}/{{ACM International Symposium}}
  on {{Microarchitecture}}}.\hskip 1em plus 0.5em minus 0.4em\relax Cambridge
  Massachusetts: ACM, Oct. 2017, pp. 273--287.

\bibitem{aliInMemoryLowCostBitSerial2020}
M.~F. Ali, A.~Jaiswal, and K.~Roy, ``In-{{Memory Low-Cost Bit-Serial Addition
  Using Commodity DRAM Technology}},'' \emph{IEEE Transactions on Circuits and
  Systems I: Regular Papers}, vol.~67, no.~1, pp. 155--165, Jan. 2020.

\bibitem{ferreiraPLUToEnablingMassively2022}
J.~D. Ferreira, G.~Falcao, J.~{G{\'o}mez-Luna}, M.~Alser, L.~Orosa,
  M.~Sadrosadati, J.~S. Kim, G.~F. Oliveira, T.~Shahroodi, A.~Nori, and
  O.~Mutlu, ``{{pLUTo}}: {{Enabling Massively Parallel Computation}} in
  {{DRAM}} via {{Lookup Tables}},'' in \emph{2022 55th {{IEEE}}/{{ACM
  International Symposium}} on {{Microarchitecture}} ({{MICRO}})}, Oct. 2022,
  pp. 900--919.

\bibitem{liDRISADRAMbasedReconfigurable2017}
S.~Li, D.~Niu, K.~T. Malladi, H.~Zheng, B.~Brennan, and Y.~Xie, ``{{DRISA}}: A
  {{DRAM-based Reconfigurable In-Situ Accelerator}},'' in \emph{Proceedings of
  the 50th {{Annual IEEE}}/{{ACM International Symposium}} on
  {{Microarchitecture}}}, ser. {{MICRO-50}} '17.\hskip 1em plus 0.5em minus
  0.4em\relax New York, NY, USA: Association for Computing Machinery, Oct.
  2017, pp. 288--301.

\bibitem{dengDrAccDRAMBased2018}
Q.~Deng, L.~Jiang, Y.~Zhang, M.~Zhang, and J.~Yang, ``{{DrAcc}}: A {{DRAM}}
  based accelerator for accurate {{CNN}} inference,'' in \emph{Proceedings of
  the 55th {{Annual Design Automation Conference}}}, ser. {{DAC}} '18.\hskip
  1em plus 0.5em minus 0.4em\relax New York, NY, USA: Association for Computing
  Machinery, Jun. 2018, pp. 1--6.

\bibitem{dengLAccExploitingLookup2019}
Q.~Deng, Y.~Zhang, M.~Zhang, and J.~Yang, ``{{LAcc}}: {{Exploiting Lookup
  Table-based Fast}} and {{Accurate Vector Multiplication}} in {{DRAM-based CNN
  Accelerator}},'' in \emph{Proceedings of the 56th {{Annual Design Automation
  Conference}} 2019}, ser. {{DAC}} '19.\hskip 1em plus 0.5em minus 0.4em\relax
  New York, NY, USA: Association for Computing Machinery, Jun. 2019, pp. 1--6.

\bibitem{lenjaniFulcrumSimplifiedControl2020}
M.~Lenjani, P.~Gonzalez, E.~Sadredini, S.~Li, Y.~Xie, A.~Akel, S.~Eilert, M.~R.
  Stan, and K.~Skadron, ``Fulcrum: {{A Simplified Control}} and {{Access
  Mechanism Toward Flexible}} and {{Practical In-Situ Accelerators}},'' in
  \emph{2020 {{IEEE International Symposium}} on {{High Performance Computer
  Architecture}} ({{HPCA}})}, Feb. 2020, pp. 556--569.

\bibitem{hajinazarSIMDRAMFrameworkBitserial2021}
N.~Hajinazar, G.~F. Oliveira, S.~Gregorio, J.~D. Ferreira, N.~M. Ghiasi,
  M.~Patel, M.~Alser, S.~Ghose, J.~{G{\'o}mez-Luna}, and O.~Mutlu,
  ``{{SIMDRAM}}: A framework for bit-serial {{SIMD}} processing using
  {{DRAM}},'' in \emph{Proceedings of the 26th {{ACM International Conference}}
  on {{Architectural Support}} for {{Programming Languages}} and {{Operating
  Systems}}}.\hskip 1em plus 0.5em minus 0.4em\relax Virtual USA: ACM, Apr.
  2021, pp. 329--345.

\bibitem{oliveiraMIMDRAMEndtoEndProcessingUsingDRAM2024}
G.~F. Oliveira, A.~Olgun, A.~G. Ya{\u g}l{\i}k{\c c}{\i}, F.~N. Bostanc{\i},
  J.~{G{\'o}mez-Luna}, S.~Ghose, and O.~Mutlu, ``{{MIMDRAM}}: {{An End-to-End
  Processing-Using-DRAM System}} for {{High-Throughput}}, {{Energy-Efficient}}
  and {{Programmer-Transparent Multiple-Instruction Multiple-Data
  Computing}},'' in \emph{2024 {{IEEE International Symposium}} on
  {{High-Performance Computer Architecture}} ({{HPCA}})}, Mar. 2024, pp.
  186--203.

\bibitem{gaoComputeDRAMInMemoryCompute2019}
F.~Gao, G.~Tziantzioulis, and D.~Wentzlaff, ``{{ComputeDRAM}}: {{In-Memory
  Compute Using Off-the-Shelf DRAMs}},'' in \emph{Proceedings of the 52nd
  {{Annual IEEE}}/{{ACM International Symposium}} on
  {{Microarchitecture}}}.\hskip 1em plus 0.5em minus 0.4em\relax Columbus OH
  USA: ACM, Oct. 2019, pp. 100--113.

\bibitem{olgunQUACTRNGHighThroughputTrue2021}
A.~Olgun, M.~Patel, A.~G. Ya{\u g}l{\i}k{\c c}{\i}, H.~Luo, J.~S. Kim,
  N.~Bostanc{\i}, N.~Vijaykumar, O.~Ergin, and O.~Mutlu, ``{{QUAC-TRNG}}:
  {{High-Throughput True Random Number Generation Using Quadruple Row
  Activation}} in {{Commodity DRAM Chips}},'' May 2021.

\bibitem{gaoFracDRAMFractionalValues2022}
F.~Gao, G.~Tziantzioulis, and D.~Wentzlaff, ``{{FracDRAM}}: {{Fractional
  Values}} in {{Off-the-Shelf DRAM}},'' in \emph{2022 55th {{IEEE}}/{{ACM
  International Symposium}} on {{Microarchitecture}} ({{MICRO}})}.\hskip 1em
  plus 0.5em minus 0.4em\relax Chicago, IL, USA: IEEE, Oct. 2022, pp. 885--899.

\bibitem{yukselFunctionallyCompleteBooleanLogic2024}
{\.I}.~E. Y{\"u}ksel, Y.~C. Tu{\u g}rul, A.~Olgun, F.~N. Bostanc{\i}, A.~G.
  Ya{\u g}l{\i}k{\c c}{\i}, G.~F. Oliveira, H.~Luo, J.~{G{\'o}mez-Luna},
  M.~Sadrosadati, and O.~Mutlu, ``Functionally-{{Complete Boolean Logic}} in
  {{Real DRAM Chips}}: {{Experimental Characterization}} and {{Analysis}},'' in
  \emph{2024 {{IEEE International Symposium}} on {{High-Performance Computer
  Architecture}} ({{HPCA}})}.\hskip 1em plus 0.5em minus 0.4em\relax Edinburgh,
  United Kingdom: IEEE, Mar. 2024, pp. 280--296.

\bibitem{yukselSimultaneousManyRowActivation2024}
{\.I}.~E. Y{\"u}ksel, Y.~C. Tu{\u g}rul, F.~N. Bostanc{\i}, G.~F. Oliveira,
  A.~G. Ya{\u g}l{\i}k{\c c}{\i}, A.~Olgun, M.~Soysal, H.~Luo,
  J.~{G{\'o}mez-Luna}, M.~Sadrosadati, and O.~Mutlu, ``Simultaneous {{Many-Row
  Activation}} in {{Off-the-Shelf DRAM Chips}}: {{Experimental
  Characterization}} and {{Analysis}},'' in \emph{2024 54th {{Annual
  IEEE}}/{{IFIP International Conference}} on {{Dependable Systems}} and
  {{Networks}} ({{DSN}})}, Jun. 2024, pp. 99--114.

\bibitem{deepseek-aiDeepSeekR1IncentivizingReasoning2025}
{DeepSeek-AI}, D.~Guo, D.~Yang, H.~Zhang, J.~Song, R.~Zhang, R.~Xu, Q.~Zhu,
  S.~Ma, P.~Wang, X.~Bi, X.~Zhang, X.~Yu, Y.~Wu, Z.~F. Wu, Z.~Gou, Z.~Shao,
  Z.~Li, Z.~Gao, A.~Liu, B.~Xue, B.~Wang, B.~Wu, B.~Feng, C.~Lu, C.~Zhao,
  C.~Deng, C.~Zhang, C.~Ruan, D.~Dai, D.~Chen, D.~Ji, E.~Li, F.~Lin, F.~Dai,
  F.~Luo, G.~Hao, G.~Chen, G.~Li, H.~Zhang, H.~Bao, H.~Xu, H.~Wang, H.~Ding,
  H.~Xin, H.~Gao, H.~Qu, H.~Li, J.~Guo, J.~Li, J.~Wang, J.~Chen, J.~Yuan,
  J.~Qiu, J.~Li, J.~L. Cai, J.~Ni, J.~Liang, J.~Chen, K.~Dong, K.~Hu, K.~Gao,
  K.~Guan, K.~Huang, K.~Yu, L.~Wang, L.~Zhang, L.~Zhao, L.~Wang, L.~Zhang,
  L.~Xu, L.~Xia, M.~Zhang, M.~Zhang, M.~Tang, M.~Li, M.~Wang, M.~Li, N.~Tian,
  P.~Huang, P.~Zhang, Q.~Wang, Q.~Chen, Q.~Du, R.~Ge, R.~Zhang, R.~Pan,
  R.~Wang, R.~J. Chen, R.~L. Jin, R.~Chen, S.~Lu, S.~Zhou, S.~Chen, S.~Ye,
  S.~Wang, S.~Yu, S.~Zhou, S.~Pan, S.~S. Li, S.~Zhou, S.~Wu, S.~Ye, T.~Yun,
  T.~Pei, T.~Sun, T.~Wang, W.~Zeng, W.~Zhao, W.~Liu, W.~Liang, W.~Gao, W.~Yu,
  W.~Zhang, W.~L. Xiao, W.~An, X.~Liu, X.~Wang, X.~Chen, X.~Nie, X.~Cheng,
  X.~Liu, X.~Xie, X.~Liu, X.~Yang, X.~Li, X.~Su, X.~Lin, X.~Q. Li, X.~Jin,
  X.~Shen, X.~Chen, X.~Sun, X.~Wang, X.~Song, X.~Zhou, X.~Wang, X.~Shan, Y.~K.
  Li, Y.~Q. Wang, Y.~X. Wei, Y.~Zhang, Y.~Xu, Y.~Li, Y.~Zhao, Y.~Sun, Y.~Wang,
  Y.~Yu, Y.~Zhang, Y.~Shi, Y.~Xiong, Y.~He, Y.~Piao, Y.~Wang, Y.~Tan, Y.~Ma,
  Y.~Liu, Y.~Guo, Y.~Ou, Y.~Wang, Y.~Gong, Y.~Zou, Y.~He, Y.~Xiong, Y.~Luo,
  Y.~You, Y.~Liu, Y.~Zhou, Y.~X. Zhu, Y.~Xu, Y.~Huang, Y.~Li, Y.~Zheng, Y.~Zhu,
  Y.~Ma, Y.~Tang, Y.~Zha, Y.~Yan, Z.~Z. Ren, Z.~Ren, Z.~Sha, Z.~Fu, Z.~Xu,
  Z.~Xie, Z.~Zhang, Z.~Hao, Z.~Ma, Z.~Yan, Z.~Wu, Z.~Gu, Z.~Zhu, Z.~Liu, Z.~Li,
  Z.~Xie, Z.~Song, Z.~Pan, Z.~Huang, Z.~Xu, Z.~Zhang, and Z.~Zhang,
  ``{{DeepSeek-R1}}: {{Incentivizing Reasoning Capability}} in {{LLMs}} via
  {{Reinforcement Learning}},'' Jan. 2025.

\bibitem{weiTMACCPURenaissance2024}
J.~Wei, S.~Cao, T.~Cao, L.~Ma, L.~Wang, Y.~Zhang, and M.~Yang, ``T-{{MAC}}:
  {{CPU Renaissance}} via {{Table Lookup}} for {{Low-Bit LLM Deployment}} on
  {{Edge}},'' Jun. 2024.

\bibitem{parkLUTGEMMQuantizedMatrix2024}
G.~Park, B.~Park, M.~Kim, S.~Lee, J.~Kim, B.~Kwon, S.~J. Kwon, B.~Kim, Y.~Lee,
  and D.~Lee, ``{{LUT-GEMM}}: {{Quantized Matrix Multiplication}} based on
  {{LUTs}} for {{Efficient Inference}} in {{Large-Scale Generative Language
  Models}},'' Apr. 2024.

\bibitem{moLUTTensorCore2024}
Z.~Mo, L.~Wang, J.~Wei, Z.~Zeng, S.~Cao, L.~Ma, N.~Jing, T.~Cao, J.~Xue,
  F.~Yang, and M.~Yang, ``{{LUT Tensor Core}}: {{Lookup Table Enables Efficient
  Low-Bit LLM Inference Acceleration}},'' Aug. 2024.

\bibitem{jedecDDR42012}
\BIBentryALTinterwordspacing
{JEDEC Solid State Technology Assn.}, ``{{DDR4 SDRAM STANDARD}},'' July 2021,
  accessed: 2024-10-30. [Online]. Available:
  \url{https://www.jedec.org/standards-documents/docs/jesd79-4a}
\BIBentrySTDinterwordspacing

\bibitem{kuboBulkBitwiseAccumulation2024}
T.~Kubo, M.~Usui, T.~Nagatani, D.~Tokuda, L.~Qu, T.~Cao, and
  S.~{Takamaeda-Yamazaki}, ``Bulk {{Bitwise Accumulation}} in {{Commercial
  DRAM}},'' in \emph{{{NeurIPS}} 2024 {{Workshop Machine Learning}} with New
  {{Compute Paradigms}}}, Oct. 2024.

\bibitem{hennessy2011computer}
J.~L. Hennessy and D.~A. Patterson, \emph{Computer architecture: a quantitative
  approach}.\hskip 1em plus 0.5em minus 0.4em\relax Elsevier, 2011.

\bibitem{gaoSeerAttentionLearningIntrinsic2025}
Y.~Gao, Z.~Zeng, D.~Du, S.~Cao, P.~Zhou, J.~Qi, J.~Lai, H.~K.-H. So, T.~Cao,
  F.~Yang, and M.~Yang, ``{{SeerAttention}}: {{Learning Intrinsic Sparse
  Attention}} in {{Your LLMs}},'' Feb. 2025.

\bibitem{wangQSparseAllLarge2024}
H.~Wang, S.~Ma, R.~Wang, and F.~Wei, ``Q-{{Sparse}}: {{All Large Language
  Models}} can be {{Fully Sparsely-Activated}},'' Jul. 2024.

\bibitem{liuTrainingFreeActivationSparsity2025}
J.~Liu, P.~Ponnusamy, T.~Cai, H.~Guo, Y.~Kim, and B.~Athiwaratkun,
  ``Training-{{Free Activation Sparsity}} in {{Large Language Models}},'' Feb.
  2025.

\bibitem{olgunDRAMBenderExtensible2023}
A.~Olgun, H.~Hassan, A.~G. Ya{\u g}l{\i}k{\c c}{\i}, Y.~C. Tu{\u g}rul,
  L.~Orosa, H.~Luo, M.~Patel, O.~Ergin, and O.~Mutlu, ``{{DRAM Bender}}: {{An
  Extensible}} and {{Versatile FPGA-based Infrastructure}} to {{Easily Test
  State-of-the-art DRAM Chips}},'' Sep. 2023.

\bibitem{yu65nm8TSRAM2022}
C.~Yu, T.~Yoo, K.~T.~C. Chai, T.~T.-H. Kim, and B.~Kim, ``A 65-nm {{8T SRAM
  Compute-in-Memory Macro With Column ADCs}} for {{Processing Neural
  Networks}},'' \emph{IEEE Journal of Solid-State Circuits}, vol.~57, no.~11,
  pp. 3466--3476, Nov. 2022.

\bibitem{ggml}
\BIBentryALTinterwordspacing
{ggml.ai}, ``{ggml},'' March 2025, accessed: 2025-03-29. [Online]. Available:
  \url{https://github.com/ggml-org/ggml}
\BIBentrySTDinterwordspacing

\bibitem{touvronLlama2Open2023}
H.~Touvron, L.~Martin, K.~Stone, P.~Albert, A.~Almahairi, Y.~Babaei,
  N.~Bashlykov, S.~Batra, P.~Bhargava, S.~Bhosale, D.~Bikel, L.~Blecher, C.~C.
  Ferrer, M.~Chen, G.~Cucurull, D.~Esiobu, J.~Fernandes, J.~Fu, W.~Fu,
  B.~Fuller, C.~Gao, V.~Goswami, N.~Goyal, A.~Hartshorn, S.~Hosseini, R.~Hou,
  H.~Inan, M.~Kardas, V.~Kerkez, M.~Khabsa, I.~Kloumann, A.~Korenev, P.~S.
  Koura, M.-A. Lachaux, T.~Lavril, J.~Lee, D.~Liskovich, Y.~Lu, Y.~Mao,
  X.~Martinet, T.~Mihaylov, P.~Mishra, I.~Molybog, Y.~Nie, A.~Poulton,
  J.~Reizenstein, R.~Rungta, K.~Saladi, A.~Schelten, R.~Silva, E.~M. Smith,
  R.~Subramanian, X.~E. Tan, B.~Tang, R.~Taylor, A.~Williams, J.~X. Kuan,
  P.~Xu, Z.~Yan, I.~Zarov, Y.~Zhang, A.~Fan, M.~Kambadur, S.~Narang,
  A.~Rodriguez, R.~Stojnic, S.~Edunov, and T.~Scialom, ``Llama 2: {{Open
  Foundation}} and {{Fine-Tuned Chat Models}},'' Jul. 2023.

\bibitem{llama3.1405B}
\BIBentryALTinterwordspacing
M.~AI, ``Introducing llama 3.1: Our most capable models to date,'' July 2024,
  accessed: 2025-03-29. [Online]. Available:
  \url{https://ai.meta.com/blog/meta-llama-3-1/}
\BIBentrySTDinterwordspacing

\bibitem{abdinPhi4TechnicalReport2024}
M.~Abdin, J.~Aneja, H.~Behl, S.~Bubeck, R.~Eldan, S.~Gunasekar, M.~Harrison,
  R.~J. Hewett, M.~Javaheripi, P.~Kauffmann, J.~R. Lee, Y.~T. Lee, Y.~Li,
  W.~Liu, C.~C.~T. Mendes, A.~Nguyen, E.~Price, G.~de~Rosa, O.~Saarikivi,
  A.~Salim, S.~Shah, X.~Wang, R.~Ward, Y.~Wu, D.~Yu, C.~Zhang, and Y.~Zhang,
  ``Phi-4 {{Technical Report}},'' Dec. 2024.

\bibitem{llama_cpp}
\BIBentryALTinterwordspacing
{ggml.ai}, ``{llama.cpp},'' March 2025, accessed: 2025-03-29. [Online].
  Available: \url{https://github.com/ggml-org/llama.cpp}
\BIBentrySTDinterwordspacing

\bibitem{hahnelMeasuringEnergyConsumption2012}
M.~H{\"a}hnel, B.~D{\"o}bel, M.~V{\"o}lp, and H.~H{\"a}rtig, ``Measuring energy
  consumption for short code paths using {{RAPL}},'' \emph{SIGMETRICS Perform.
  Eval. Rev.}, vol.~40, no.~3, pp. 13--17, Jan. 2012.

\bibitem{tegrastats}
\BIBentryALTinterwordspacing
{NVIDIA}, ``{tegrastats Utility},'' November 2023, accessed: 2025-03-29.
  [Online]. Available:
  \url{https://developer.nvidia.com/docs/drive/drive-os/6.0.8.1/public/drive-os-linux-sdk/common/topics/util_setup/tegrastatsUtility1.html}
\BIBentrySTDinterwordspacing

\bibitem{balasubramonianCACTI7New2017}
R.~Balasubramonian, A.~B. Kahng, N.~Muralimanohar, A.~Shafiee, and V.~Srinivas,
  ``{{CACTI}} 7: {{New Tools}} for {{Interconnect Exploration}} in {{Innovative
  Off-Chip Memories}},'' \emph{ACM Trans. Archit. Code Optim.}, vol.~14, no.~2,
  pp. 14:1--14:25, Jun. 2017.

\bibitem{namDRAMScopeUncoveringDRAM2024a}
H.~Nam, S.~Baek, M.~Wi, M.~J. Kim, J.~Park, C.~Song, N.~S. Kim, and J.~H. Ahn,
  ``{{DRAMScope}}: {{Uncovering DRAM Microarchitecture}} and
  {{Characteristics}} by {{Issuing Memory Commands}},'' in \emph{2024
  {{ACM}}/{{IEEE}} 51st {{Annual International Symposium}} on {{Computer
  Architecture}} ({{ISCA}})}, Jun. 2024, pp. 1097--1111.

\bibitem{marazziHiFiDRAMEnablingHighfidelity2024}
M.~Marazzi, T.~Sachsenweger, F.~Solt, P.~Zeng, K.~Takashi, M.~Yarema, and
  K.~Razavi, ``{{HiFi-DRAM}}: {{Enabling High-fidelity DRAM Research}} by
  {{Uncovering Sense Amplifiers}} with {{IC Imaging}},'' in \emph{2024
  {{ACM}}/{{IEEE}} 51st {{Annual International Symposium}} on {{Computer
  Architecture}} ({{ISCA}})}, Jun. 2024, pp. 133--149.

\bibitem{hoonProcessinginMemoryArchitectureCommodity2024}
S.~Hoon, ``A {{Processing-in-Memory Architecture}} for {{Commodity DRAM
  Devices}} with {{Enhanced Reliability}} and {{Compatibility}},'' Feb. 2024.

\bibitem{leeHardwareArchitectureSoftware2021}
S.~Lee, S.-h. Kang, J.~Lee, H.~Kim, E.~Lee, S.~Seo, H.~Yoon, S.~Lee, K.~Lim,
  H.~Shin, J.~Kim, O.~Seongil, A.~Iyer, D.~Wang, K.~Sohn, and N.~S. Kim,
  ``Hardware {{Architecture}} and {{Software Stack}} for {{PIM Based}} on
  {{Commercial DRAM Technology}} : {{Industrial Product}},'' in \emph{2021
  {{ACM}}/{{IEEE}} 48th {{Annual International Symposium}} on {{Computer
  Architecture}} ({{ISCA}})}.\hskip 1em plus 0.5em minus 0.4em\relax Valencia,
  Spain: IEEE, Jun. 2021, pp. 43--56.

\bibitem{kwonSystemArchitectureSoftware2022}
Y.~Kwon, K.~Vladimir, N.~Kim, W.~Shin, J.~Won, M.~Lee, H.~Joo, H.~Choi, G.~Kim,
  B.~An, J.~Kim, J.~Lee, I.~Kim, J.~Park, C.~Park, Y.~Song, B.~Yang, H.~Lee,
  S.~Kim, D.~Kwon, S.~Lee, K.~Kim, S.~Oh, J.~Park, G.~Hong, D.~Ka, K.~Hwang,
  J.~Park, K.~Kang, J.~Kim, J.~Jeon, M.~Lee, M.~Shin, M.~Shin, J.~Cha, C.~Jung,
  K.~Chang, C.~Jeong, E.~Lim, I.~Park, J.~Chun, and S.~Hynix, ``System
  {{Architecture}} and {{Software Stack}} for {{GDDR6-AiM}},'' in \emph{2022
  {{IEEE Hot Chips}} 34 {{Symposium}} ({{HCS}})}, Aug. 2022, pp. 1--25.

\bibitem{devauxTrueProcessingMemory2019}
F.~Devaux, ``The true {{Processing In Memory}} accelerator,'' in \emph{2019
  {{IEEE Hot Chips}} 31 {{Symposium}} ({{HCS}})}, Aug. 2019, pp. 1--24.

\bibitem{kwon25420nm6GB2021}
Y.-C. Kwon, S.~H. Lee, J.~Lee, S.-H. Kwon, J.~M. Ryu, J.-P. Son, O.~Seongil,
  H.-S. Yu, H.~Lee, S.~Y. Kim, Y.~Cho, J.~G. Kim, J.~Choi, H.-S. Shin, J.~Kim,
  B.~Phuah, H.~Kim, M.~J. Song, A.~Choi, D.~Kim, S.~Kim, E.-B. Kim, D.~Wang,
  S.~Kang, Y.~Ro, S.~Seo, J.~Song, J.~Youn, K.~Sohn, and N.~S. Kim, ``25.4
  {{A}} 20nm {{6GB Function-In-Memory DRAM}}, {{Based}} on {{HBM2}} with a
  1.{{2TFLOPS Programmable Computing Unit Using Bank-Level Parallelism}}, for
  {{Machine Learning Applications}},'' in \emph{2021 {{IEEE International
  Solid-State Circuits Conference}} ({{ISSCC}})}, vol.~64, Feb. 2021, pp.
  350--352.

\bibitem{liSpecPIMAcceleratingSpeculative2024}
C.~Li, Z.~Zhou, S.~Zheng, J.~Zhang, Y.~Liang, and G.~Sun, ``{{SpecPIM}}:
  {{Accelerating Speculative Inference}} on {{PIM-Enabled System}} via
  {{Architecture-Dataflow Co-Exploration}},'' in \emph{Proceedings of the 29th
  {{ACM International Conference}} on {{Architectural Support}} for
  {{Programming Languages}} and {{Operating Systems}}, {{Volume}} 3}.\hskip 1em
  plus 0.5em minus 0.4em\relax La Jolla CA USA: ACM, Apr. 2024, pp. 950--965.

\end{thebibliography}

\end{document}